\pdfoutput=1
\documentclass[conference]{IEEEtran}
\IEEEoverridecommandlockouts

\usepackage[T1]{fontenc}
\usepackage[utf8]{inputenc}
\usepackage{graphicx}
\usepackage{amsmath}
\usepackage{amssymb}
\usepackage{booktabs}
\usepackage{multirow}
\usepackage{tabularx}
\usepackage{array}
\usepackage{xcolor}
\usepackage{cite}
\usepackage{url}
\usepackage{siunitx}
\usepackage{tikz}
\usetikzlibrary{positioning,fit,backgrounds,arrows.meta,calc,shapes.misc,decorations.pathreplacing}
\usepackage[hidelinks]{hyperref}
\usepackage{microtype}
\usepackage{float}

\graphicspath{{figures/}}

\newcolumntype{L}[1]{>{\raggedright\arraybackslash}p{#1}}
\newcolumntype{R}[1]{>{\raggedleft\arraybackslash}p{#1}}
\newcolumntype{Y}{>{\raggedright\arraybackslash}X}

\newcommand{\us}{\ensuremath{\mu}s}
\newcommand{\ocudu}{OCUDU}
\newcommand{\ClassA}{Class~A}
\newcommand{\ClassB}{Class~B}
\newcommand{\ClassC}{Class~C}

\newcommand{\etal}{et~al.}
\newcommand{\code}[1]{\texttt{#1}}

\definecolor{clsA}{RGB}{170,40,40}
\definecolor{clsB}{RGB}{40,120,50}
\definecolor{clsC}{RGB}{20,110,120}

\begin{document}

\title{Real-Time dApps for AI-RAN: Measured Interface Requirements for Inline PHY and Slot-Level Control}

\author{
\IEEEauthorblockN{Timothy O'Shea, Matthew Pennybacker, Andriy Kharchenko}
\IEEEauthorblockA{DeepSig Inc., Arlington, VA, USA}
}

\maketitle
\begingroup
\renewcommand{\thefootnote}{\fnsymbol{footnote}}
\footnotetext[1]{Preview version. This paper accompanies a preview release of the platform and its benchmark suite and will receive updates as the working group's review, new use cases, and measurements land.}
\endgroup

\newcommand{\ChkARxFull}{272\,\us{} / 523\,\us{}}
\newcommand{\ChkARxFullSubmit}{45.9\,\us{}}
\newcommand{\ChkARxLive}{81.6\,\us{} / 112\,\us{}}
\newcommand{\ChkAStream}{19.9\,\us{} / 27.7\,\us{}}
\newcommand{\ChkAStreamFifty}{19.9\,\us{}}
\newcommand{\ChkBAbi}{0.288\,\us{}}
\newcommand{\ChkBCtx}{48\,ns / 0.27\,\us{}}
\newcommand{\ChkCERT}{13.0\,\us{} / 51.7\,\us{}}
\newcommand{\ChkCERTsuite}{10.9\,\us{} / 11.9\,\us{} (RTT/2)}
\newcommand{\ChkCIdle}{0.032\,\us{} / 0.048\,\us{}}
\newcommand{\ChkCRing}{2.05\,\us{} / 3.89\,\us{}}
\newcommand{\ConflateAge}{130\,\us{}}
\newcommand{\ConflateDrops}{10588}
\newcommand{\ContBAbi}{5.22\,\us{}}
\newcommand{\ContBAbiMiss}{0.005\,\%}
\newcommand{\ContBSctp}{181\,\us{}}
\newcommand{\ContBSctpMiss}{0.35\,\%}
\newcommand{\ContBShm}{16.8\,\us{}}
\newcommand{\ContBShmMiss}{0.04\,\%}
\newcommand{\ContBZmq}{787\,\us{}}
\newcommand{\ContBZmqMiss}{0.69\,\%}
\newcommand{\ContEOneWaySmall}{216\,\us{}}
\newcommand{\ContEOneWaySrs}{1.57\,ms}
\newcommand{\ContERT}{492\,\us{}}
\newcommand{\ContERTFifty}{12.3\,\us{}}
\newcommand{\ContRingProd}{167\,\us{}}
\newcommand{\ContRingWake}{1.62\,ms}
\newcommand{\DrvAbiRT}{0.288\,\us{}}
\newcommand{\DrvDtoH}{73.5\,\us{}}
\newcommand{\DrvEGrid}{1.68\,ms}
\newcommand{\DrvEInd}{1.43\,ms}
\newcommand{\DrvESrs}{255\,\us{}}
\newcommand{\DrvRingProd}{2.34\,\us{}}
\newcommand{\DrvSctpRT}{15.6\,\us{}}
\newcommand{\EThreeCtlDec}{2.29\,\us{}}
\newcommand{\EThreeCtlEnc}{2.27\,\us{}}
\newcommand{\EThreeDecRatio}{0.97$\times$}
\newcommand{\EThreeFbGrid}{42.1\,\us{} to encode and 11.3\,\us{} to decode}
\newcommand{\EThreeGridEnc}{681\,\us{}}
\newcommand{\EThreeOneWayBig}{1.68\,ms}
\newcommand{\EThreeOneWayGrid}{1.43\,ms}
\newcommand{\EThreeOneWayGridTail}{1.97\,ms}
\newcommand{\EThreeOneWaySmall}{10.5\,\us{}}
\newcommand{\EThreeOneWaySrs}{255\,\us{}}
\newcommand{\EThreePyRT}{1.39\,ms one way (fixture timer bound)}
\newcommand{\EThreeRT}{12.6\,\us{}}
\newcommand{\EThreeRTFifty}{10.9\,\us{}}
\newcommand{\EThreeSlope}{0.87\,ns}
\newcommand{\EThreeSrsEnc}{57.8\,\us{}}
\newcommand{\JbpfHookGrid}{95.4\,\us{}}
\newcommand{\JbpfHookMid}{2.61\,\us{}}
\newcommand{\JbpfHookSmall}{1.54\,\us{}}
\newcommand{\JbpfRT}{152\,\us{}}
\newcommand{\LeaseProd}{0.032\,\us{}}
\newcommand{\MeasAText}{The \ClassA{} rows are inexpressible on an external boundary for the structural reason of Section~\ref{sec:corpus}, so the measurement answers a narrower question: were a framework extended with a return path, what would the export alone cost before any inference ran? On the coherent-memory host a device-to-host copy of a 1\,MB grid into pinned memory takes \MechADtoHFifty{} at the median and \MechADtoH{} at P99. The full NVIDIA-style path (copy, shared-memory stage, ZeroMQ notification) takes \MechANvFifty{} at the median and \MechANvNN{} at P99 for the 1.47\,MB four-port grid, against a 300\,\us{} planning budget. Both paths measure the outbound leg alone. A stage needs the result back, so the round trip is at least twice the outbound figure, about 450\,\us{} at the median for the NVIDIA-style path before a single model FLOP, and no framework provides the return copy. The in-process figure is the cost of enqueueing the dApp's kernels on the lane's stream and recording the completion event, measured with the released packages on the exclusive GPU. At the live shape (51\,PRB, two layers, 256-QAM), the channel estimator submits in \ChkAStreamFifty{} at the median and the reference receiver completes in \ChkARxLive{} P50 / P99.9, inside its 150\,\us{} qualification deadline. At the 273-PRB envelope the same receiver submits in \ChkARxFullSubmit{} but completes in \ChkARxFull{}, so the reference kernels, not the interface, are what remains to be qualified at that shape. The framework-path suite shows the same thing at every payload from 256\,KB to 23\,MB: residency costs a launch, export costs a copy proportional to the tensor, and at the 64-port envelope the copy alone (\MechASixtyFour{} at 23.5\,MB) exceeds the slot.
}
\newcommand{\MeasBText}{The \ClassB{} question is whether a process boundary can sit inside a 100\,\us{} decision. All six paths carry the same request, a 16-candidate scheduler input of 3{,}240\,B (the released \code{scheduler\_input\_v1} with two 275-bit masks) returning 16 intents of 72\,B. Each path invokes the same dApp function through the released interface's \code{invoke} pointer: a direct call; a heap SPSC pair between two pinned threads; a shared-memory SPSC across a fork; ZeroMQ request/reply over \code{ipc://} and again over TCP loopback; and an SCTP one-to-one association on loopback with real message boundaries, the carrier an E3 control action would use. The direct call is timed with the DU's admission checks included, which cover structure size and ABI major, candidate identity, allow flags, PRB range, MCS bound, and deadline. Each path runs 20{,}000 requests on separate pinned cores under three conditions: the quiet host, the quiet host with a 50\,\us{}-on, 50\,\us{}-off busy thread on the responder's core, and the host with the DU on the air.

Fig.~\ref{fig:classb} shows the result. On the quiet host every carrier meets the deadline: the direct call completes in \MechBAbiFifty{} at the median and \MechBAbi{} at P99.9 with validation included, the shared-memory SPSC in \MechBShmFifty{} and \MechBShm{}, SCTP in \MechBSctpFifty{} and \MechBSctp{}, ZeroMQ over \code{ipc://} in \MechBZmqFifty{} and \MechBZmq{}. The antagonist adds about 50\,\us{} to every tail (\MechBSctpAnt{} for SCTP, \MechBZmqAnt{} for ZeroMQ, \MechBAbiAnt{} for the direct call, whose caller shared the contested core); all still fit. With the DU on the air the picture changes: the direct call moves to \ContBAbi{} at P99.9 and the SPSC to \ContBShm{}, while the message paths move to \ContBSctp{} (SCTP) and \ContBZmq{} (ZeroMQ) and miss the deadline on \ContBSctpMiss{} and \ContBZmqMiss{} of requests with the responder core otherwise idle. APER encode and decode at both ends of the SCTP path add up to \MechBE{} at the median on the quiet host, a floor for a compiled E3 control loop; the published loops of about 400\,\us{}~\cite{lacava2025dapps} include a Python decoder and the agent's data plane, and report no tails.

The reading is not that ZeroMQ or SCTP are slow; on an idle host they would do. It is that a bounded contract is defined by its tail under the load the DU itself creates, and the tail of any path with a wake in it is then set by the kernel and by whatever else the DU runs. Nor is a miss free because the contract has a fallback. The conventional decision is taken at the deadline, but the MAC has already spent the wait, and a \ContBZmq{} tail is longer than the slot: the request outlives the decision it was for, the reply must be discarded on arrival, and the intents for the next slot are computed from a context one slot stale. A miss rate of \ContBSctpMiss{} at three UEs on a 51-PRB cell is the floor of that cost, not its ceiling, since Boeira~\etal~\cite{boeira2026perf} show the same framework's CPU saturating before its GPU as dApps are added.

The busy-polling shared-memory SPSC deserves its own reading, because it is a process boundary that stays inside the budget under load, \ContBShm{} at P99.9 with \ContBShmMiss{} misses. It is the supervised \ClassB{} position the spectrum admits: one copy of 3\,KB, one copy of 1\,KB back, crash isolation, and no wake, at the price of a core the responder must spin on and the DU therefore loses. The platform defines that position as an alternative and does not ship it, for two reasons the measurements make concrete. The validation a \ClassB{} call needs is the same on either side of the boundary, 72-byte intents range-checked field by field, so isolation buys protection against a crash but not against a wrong answer, and the circuit breaker already bounds the crash. And the direct call is still an order of magnitude tighter, \ContBAbi{} against \ContBShm{}, on a budget that the feature block and commit also have to share. The one path whose tail stays inside the budget on the host that matters without reserving a core is the direct call; the price, validating 72\,B intents, is small (Section~\ref{sec:design}).
}
\newcommand{\MeasCText}{A \ClassC{} publisher runs on the PHY thread. Whatever the consumer does, the producer's cost per slot must stay bounded and small, and the data the consumer eventually sees must have a known age. Fig.~\ref{fig:producer} tests both properties for the \ocudu{} positions and for the ZeroMQ publish/subscribe path of the observer frameworks, at the live 68\,KB slot.

The supervised ring is the released layout in a two-process harness: try-lock guard, reclaim read, memcpy into the next 64\,B-aligned slot, release store, eventfd write, \RingSyscalls{} system calls per publication. Its critical section costs \RingProdFifty{} at the median for 68\,KB and \RingProdBig{} for the 1.47\,MB envelope, which is the memcpy; when the ring is full the producer returns in \RingProdFull{} with a counted drop and never waits. In the slot-timed run the producer publishes every 500\,\us{} for \SlotN{} slots while the consumer either drains promptly or sleeps 5\,ms on every 8th slot. With the sleeping consumer the ring drops \SlotDrops{} slots across \SlotStalls{} stalls, about two per stall, which is exactly the excess of a 5\,ms stall over an 8-slot ring, and the age of the newest data the consumer sees is bounded at \SlotRingAge{} median and \SlotRingAgeMax{} P99.9 by the ring depth. An unsubscribed stream costs the producer one relaxed atomic load, \SlotIdle{}.

ZeroMQ over \code{ipc://} publishes the same 68\,KB in \ZmqProdFifty{} at the median, the same producer cost; the difference is what happens to the data. With the default high-water mark the socket queues everything a slow subscriber has not read, so the age of the data it eventually sees grows to \ZmqAgeSlow{} at the median and \ZmqAgeSlowMax{} at P99.9, with no drop reported; with the conflating subscriber option the Python dApps use, the age stays at one slot but \ConflateDrops{} slots are discarded inside the socket with no count anywhere. The ring is not faster than the socket; it is \emph{accountable}: staleness is bounded by a configured depth and every drop is counted, which is what an operator needs to know whether an advisory loop is still advising. On the quiet host the ring producer's P99.9 is \RingProd{}; with the DU on the air it is \ContRingProd{}, and the consumer's wake tail grows from \RingWake{} to \ContRingWake{} (Table~\ref{tab:conditions}), because the bench cores then also carry DU worker threads and enter deep idle states between slots.

The SEQPACKET control channel, credentials verified on every receive, completes a 256\,B request/response in \SeqRT{} at the median and \SeqRTTail{} at P99.9 and a heartbeat in \SeqHb{}, against \SeqZmq{} for ZeroMQ request/reply: the cost of noticing a dead worker within one heartbeat and restarting it without touching the ring. The consumer's wake from eventfd to a leased slot is \RingWakeFifty{} at the median and \RingWake{} at P99.9. The native lease costs the producer only the unsubscribed check.
}
\newcommand{\MeasEText}{Fig.~\ref{fig:e3} gives the cost of the protocol-standard end with nothing modeled. The codec rows encode and decode real E3AP PDUs with the \ocudu{} asn1c APER codec: setup, subscription, indications with bodies from 64\,B to a fragmented 734\,KB grid, a 2.2\,KB scheduler control action, and an acknowledgment. Two regimes appear. For control-sized messages the cost is the envelope, \EThreeCtlEnc{} to encode and \EThreeCtlDec{} to decode a control action, independent of the few hundred bytes of body, and the Northeastern codec tracks it closely. For indications the cost grows linearly at \EThreeSlope{} per byte because APER copies an \code{OCTET STRING} body bit-aligned: a 64\,KB SRS indication costs \EThreeSrsEnc{} and a 734\,KB grid \EThreeGridEnc{} to encode, with decode near \EThreeDecRatio{} the encode cost. The FlatBuffers data profile on the second association costs \EThreeFbGrid{} for the same grid, which is why the platform has two associations.

The carrier rows send those PDUs between two pinned processes over an SCTP one-to-one association on loopback, one PDU per message, fragmented at 64\,KiB as the E3AP wire contract requires, and time one-way delivery through reassembly and decode. A 1.5\,KB indication arrives in \EThreeOneWaySmall{} at the median; the 734\,KB grid, 12 fragments, in \EThreeOneWayGrid{} (\EThreeOneWayGridTail{} at P99.9); the 1.47\,MB four-port grid in \EThreeOneWayBig{}. A control round trip on stream 0 with the real codec at both ends is \EThreeRTFifty{} at the median and \EThreeRT{} at P99.9; the published loops of about 400\,\us{}~\cite{lacava2025dapps} sit between these compiled figures and the same path through a Python decoder. With the DU on the air the medians barely move and the tails do: the 1.5\,KB indication reaches \ContEOneWaySmall{} at P99.9, the 64\,KB SRS indication \ContEOneWaySrs{}, and the control round trip \ContERT{} (Table~\ref{tab:conditions}).

These numbers draw the lines on the corpus map of Fig.~\ref{fig:corpus_map}. Every \ClassC{} row lies above both, most by an order of magnitude, so an E3 client is a correct implementation of all 17. Every \ClassA{} row lies below the quiet-host line, so even before the missing return path is counted the export alone does not fit. Every \ClassB{} row lies above the quiet-host line and below the DU-on-air line: the boundary fits the 100\,\us{} budget on an idle machine and not on the machine the DU runs on. That is the whole case for the right-hand end: it costs nothing the asynchronous rows cannot afford, and no codec optimization moves the lines, because they are set by the copy, the wake, and the association.
}
\newcommand{\MeasPathsText}{Fig.~\ref{fig:paths} closes the section by putting the frameworks side by side under the harness above. For each of three workloads, one per class, the figure shows the median and P99.9 of each framework's path at the workload's byte count against its budget; published anchors from the cited papers appear as marks without percentiles. The picture is the one the corpus predicted. On the \ClassA{} workload only the resident path is inside the budget, and the other frameworks' paths are shown hatched because they are outside their authors' stated scope. On the \ClassB{} workload the resident path is more than two orders of magnitude inside the budget and the observer path is outside it. On the \ClassC{} workload all three are inside the budget by a wide margin, and the differences among them are the producer-side and restart properties measured above rather than feasibility.

\begin{figure*}[t]
\centering
\includegraphics[width=0.9\textwidth]{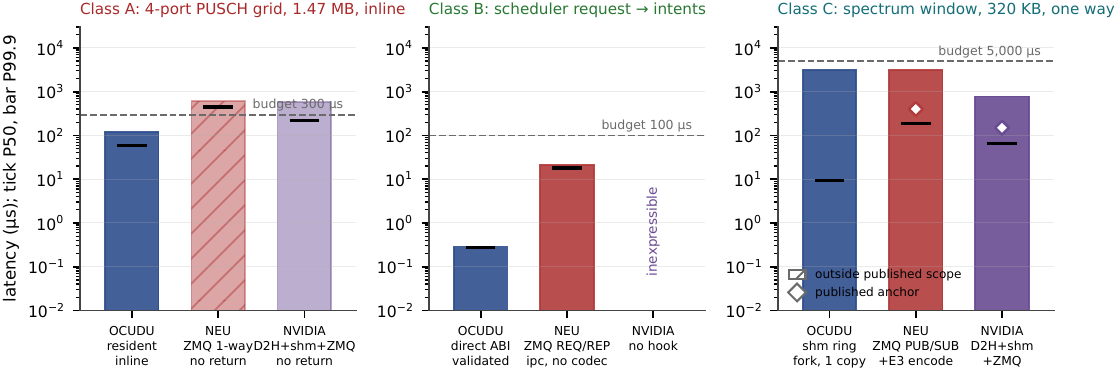}
\caption{Contract-matched framework paths, one workload per class, P50 (tick) and P99.9 (bar) against the workload budget (dashed). Hatched bars are outside the framework's published scope and are shown as sensitivity only; marks without bars are published anchors.}
\label{fig:paths}
\end{figure*}
}
\newcommand{\MechADtoH}{80.6\,\us{}}
\newcommand{\MechADtoHFifty}{73.5\,\us{}}
\newcommand{\MechANvFifty}{224\,\us{}}
\newcommand{\MechANvNN}{306\,\us{}}
\newcommand{\MechASixtyFour}{about 17.4\,ms at the measured 11\,Gb/s device-to-host rate}
\newcommand{\MechAStream}{submit 19.9\,\us{} P50, 27.7\,\us{} P99.9 (estimator, live shape)}
\newcommand{\MechBAbi}{0.288\,\us{}}
\newcommand{\MechBAbiAnt}{53.2\,\us{}}
\newcommand{\MechBAbiFifty}{0.272\,\us{}}
\newcommand{\MechBAbiMiss}{0}
\newcommand{\MechBE}{22.2\,\us{}}
\newcommand{\MechBSctp}{15.6\,\us{}}
\newcommand{\MechBSctpAnt}{65.8\,\us{}}
\newcommand{\MechBSctpFifty}{13.1\,\us{}}
\newcommand{\MechBSctpMiss}{0.005\,\%}
\newcommand{\MechBShm}{0.944\,\us{}}
\newcommand{\MechBShmAnt}{54.1\,\us{}}
\newcommand{\MechBShmFifty}{0.784\,\us{}}
\newcommand{\MechBZmq}{21.3\,\us{}}
\newcommand{\MechBZmqAnt}{73.4\,\us{}}
\newcommand{\MechBZmqFifty}{18.0\,\us{}}
\newcommand{\MechBZmqMiss}{0.005\,\%}
\newcommand{\MechBZmqMissAnt}{0.000\,\%}
\newcommand{\MechCCtx}{48\,ns P50, 0.27\,\us{} P99.9}
\newcommand{\MechCEGrid}{1.68\,ms}
\newcommand{\MechCEInd}{255\,\us{}}
\newcommand{\MechCERT}{10.9\,\us{} P50 (RTT/2)}
\newcommand{\MechCIdle}{0.048\,\us{} P99.9}
\newcommand{\MechCRingProd}{2.34\,\us{}}
\newcommand{\MechCRingSlow}{0.144\,\us{} (full ring, counted drop)}
\newcommand{\MechCSeq}{3.20\,\us{} P50 (RTT/2)}
\newcommand{\MechCWake}{3.92\,\us{} P50}
\newcommand{\MechCZmqProd}{3.14\,\us{} P99.9}
\newcommand{\MechCZmqSlow}{171\,\us{} P99.9}
\newcommand{\PosEC}{255\,\us{}}
\newcommand{\PosEP}{57.8\,\us{}}
\newcommand{\PosEPT}{520\,\us{}}
\newcommand{\PosNatC}{0 (in place)}
\newcommand{\PosNatP}{0.032\,\us{}}
\newcommand{\PosNatPT}{0.048\,\us{}}
\newcommand{\PosRingC}{3.92\,\us{}}
\newcommand{\PosRingP}{1.97\,\us{}}
\newcommand{\PosRingPT}{2.34\,\us{}}
\newcommand{\PosZmqC}{130\,\us{}}
\newcommand{\PosZmqP}{1.82\,\us{}}
\newcommand{\PosZmqPT}{3.14\,\us{}}
\newcommand{\RingProd}{2.34\,\us{}}
\newcommand{\RingProdBig}{23.9\,\us{}}
\newcommand{\RingProdFifty}{1.97\,\us{}}
\newcommand{\RingProdFull}{0.144\,\us{}}
\newcommand{\RingSyscalls}{3.0}
\newcommand{\RingWake}{99.4\,\us{}}
\newcommand{\RingWakeFifty}{3.92\,\us{}}
\newcommand{\SeqHb}{3.06\,\us{}}
\newcommand{\SeqRT}{3.20\,\us{}}
\newcommand{\SeqRTTail}{4.26\,\us{}}
\newcommand{\SeqStream}{3.06\,\us{}}
\newcommand{\SeqZmq}{8.52\,\us{}}
\newcommand{\SlotDrops}{4457}
\newcommand{\SlotIdle}{0.032\,\us{} P50, 0.048\,\us{} P99.9}
\newcommand{\SlotN}{20{,}000}
\newcommand{\SlotRingAge}{3.15\,ms}
\newcommand{\SlotRingAgeFast}{4.21\,\us{}}
\newcommand{\SlotRingAgeMax}{5.41\,ms}
\newcommand{\SlotRingFast}{3.89\,\us{}}
\newcommand{\SlotRingSlow}{2.00\,\us{}}
\newcommand{\SlotStalls}{1942}
\newcommand{\TabClassB}{Direct ABI (validated) & 0.27 & 0.29\,/\,0.29 & 0.29\,/\,53 & 0.000\,\%\,/\,0.000\,\% \\
SPSC, 2 threads & 0.75 & 0.86\,/\,0.88 & 1.14\,/\,54 & 0.000\,\%\,/\,0.000\,\% \\
SPSC shm, fork & 0.78 & 0.88\,/\,0.90 & 0.94\,/\,54 & 0.000\,\%\,/\,0.000\,\% \\
SCTP loopback & 13 & 14\,/\,63 & 16\,/\,66 & 0.005\,\%\,/\,0.000\,\% \\
ZeroMQ \code{ipc://} & 18 & 20\,/\,72 & 21\,/\,73 & 0.005\,\%\,/\,0.000\,\% \\
ZeroMQ TCP loopback & 23 & 25\,/\,76 & 27\,/\,79 & 0.005\,\%\,/\,0.01\,\% \\}
\newcommand{\TabEThree}{Setup resp.\ (2 RAN fn.) & 1231 & 2.82 & 3.02 & 9.8 & 11 \\
Subscription req. & 62 & 0.672 & 0.688 & 7.5 & 9 \\
Control action (2.2\,KB) & 2192 & 2.27 & 2.29 & 10.9 & 13 \\
Indication (1.5\,KB) & 1565 & 1.92 & 2.00 & 10.5 & 13 \\
Indication (64\,KB SRS) & 65565 & 57.8 & 56.3 & 255.1 & 520 \\
Indication (734\,KB, 12 frag.) & 752073 & 681 & 664 & 1428.9 & 1966 \\
Indication (1.47\,MB, 23 frag.) & 1506155 & -- & -- & 1681.0 & 1974 \\
FlatBuffers profile (734\,KB) & 752616 & 42 & 11.3 & -- & -- \\}
\newcommand{\ZmqAgeSlow}{981.95\,ms}
\newcommand{\ZmqAgeSlowMax}{1045.61\,ms}
\newcommand{\ZmqProdFast}{3.14\,\us{}}
\newcommand{\ZmqProdFifty}{1.82\,\us{}}
\newcommand{\ZmqProdSlow}{171\,\us{}}

\begin{abstract}
Distributed applications (dApps) bring AI to the microsecond-to-millisecond band beside the 5G distributed unit (DU), but every public dApp framework realizes them the same way: an external process that receives an indication and returns a control message. That boundary is right for sensing and advisory workloads. It cannot express a neural receiver that must finish inside a slot, and it cannot hold a scheduling decision the MAC is waiting on. This paper asks what a dApp interface must deliver, in latency and bandwidth, for the AI-RAN use cases now filed under that name to be realized as dApps at all. It treats the coupling between a dApp and the DU as a design axis with two legitimate ends, a C ABI inside the DU process and a protocol-standard E3AP association over SCTP, and places the three \ocudu{} dApp classes on it. An audited corpus of 39 runtime AI-RAN use cases, sized by 5G NR timing, shows that more than half cannot cross the observer boundary: inline PHY work because an indication has no return path into the same slot, and bounded control because of the tail under load. Measuring the mechanisms each framework actually uses, on a quiet host and with a live cell on the air, shows that every carrier meets a 100\,\us{} control deadline on an idle host and that only the in-process paths still do once the DU is running. The asynchronous use cases remain feasible at every position, so the protocol-standard end is kept as a first-class option; what in-process placement adds for them is accountable staleness and an observation-to-decision path with no message on it. These measurements derive the released ABI, a stream for inline work, a validated call for bounded control, and a choice of lease, supervised ring, or portable E3 client for observation, and four dApps of all three classes are validated together on one over-the-air cell.
\end{abstract}


\section{Introduction}
\label{sec:intro}

The O-RAN application hierarchy has two established tiers. rApps attach to the non-real-time RIC and act at management timescales of seconds and above; xApps attach to the near-real-time RIC through E2 and close control loops between 10\,ms and 1\,s~\cite{oran_wg1,polese_oran}. Neither can observe or act on the state of a distributed unit (DU) within a slot, a few slots, or a frame. Distributed applications (dApps) were proposed to fill that band~\cite{bonati2022dapps,lacava2025dapps}: they run on the node that owns the data, consume observations that never leave it, and close loops below 10\,ms.

Every dApp framework published so far realizes that idea the same way. The dApp is a separate process; the DU exports an indication, the process computes, and a control message comes back~\cite{lacava2025dapps,villa2026aerialdapp,santhi2025interforan,boeira2026perf}. This is a \emph{tap}: an observer with a reply channel. It is the right shape for spectrum sensing, interference detection, or integrated sensing and communication (ISAC), where a result 2\,ms late is still useful and a dropped one is harmless. It is the wrong shape for two other kinds of work the AI-RAN literature also files under ``dApp'': a neural channel estimator or receiver that must finish inside a PUSCH slot over tensors a GPU already holds, and a scheduling or link-adaptation decision the MAC is waiting on now. The first is not an observer at all but a \emph{stage}, a replacement for a function in the pipeline, and for it the export is the cost. The second is a \emph{bounded call}, and for it the process boundary is the risk.

The \ocudu{} platform names these three contracts as classes: \ClassA{}, resident inline L1 work that runs on the PHY's own stream; \ClassB{}, bounded real-time control that answers the MAC inside an admitted deadline; and \ClassC{}, asynchronous observation and advisory work whose producer never waits. The companion papers define the classes from a use-case corpus~\cite{pennybacker2026dappclasses} and describe the embedded runtime that implements them inside the \ocudu{} DU~\cite{pennybacker2026dapparch}. This paper answers the question they leave open: \emph{why these interface mechanisms, at these costs, and why offer more than one}. It measures the mechanisms the three public frameworks actually use, at the payload shapes the use cases produce, on the platform where the released runtime runs. Fig.~\ref{fig:paths} shows the outcome in one picture: one workload per class, each framework's own path against the workload's budget.

Three findings carry the paper. First, 22 of the 39 runtime use cases in the audited corpus cannot cross the observer boundary, 13 of them for a structural reason that no faster codec changes: an indication carries data outward, and nothing brings a tensor back into the same slot. Second, a process boundary inside a 100\,\us{} decision is fine on an idle host and not on the host that matters. Every carrier meets the \ClassB{} deadline on the quiet host; with the DU on the air, the same request over SCTP loopback reaches \ContBSctp{} at P99.9 and over ZeroMQ \ContBZmq{}, while the in-process validated call stays at \ContBAbi{}. Third, for the 17 asynchronous rows the observer boundary is the right one, and what the in-process \ClassC{} positions add is not speed but accountability, bounded staleness with counted drops, and a path by which an observation reaches a per-slot decision without a message. The paper contributes:
\begin{enumerate}
\item a \textbf{coupling spectrum} for dApp interfaces with two legitimate ends, a tightly coupled C ABI inside the DU process and a protocol-standard out-of-process carrier (E3AP over SCTP), on which the three classes are positions and \ClassC{} spans three;
\item an \textbf{audited corpus} of 39 runtime AI-RAN use cases with deadlines from 5G NR timing, byte counts at the live and envelope shapes, and a verdict on what an observer-only framework, a GPU-export framework, and an in-process ABI can each express;
\item \textbf{measurements of the real mechanisms} under one pinned harness with raw percentiles, on a quiet host and with the DU on the air: the asn1c APER E3AP codec and SCTP carrier, the \ocudu{} shared ring, SEQPACKET supervisor, and validated direct call, jBPF, and the ZeroMQ and device-to-host paths of the other frameworks;
\item a \textbf{derivation of the released ABI} from those measurements, including the case for keeping the protocol-standard end as a first-class \ClassC{} option and for the one position the spectrum admits but the platform does not ship; and
\item \textbf{validation}: four dApps of three classes composed on one over-the-air cell, with the runtime's per-class checkpoints.
\end{enumerate}
Every number in the paper is produced by the accompanying suite or by the released runtime, every mechanism named is the one the runtime ships, and the suite, corpus, and raw traces are public in the study repository~\cite{ocudu_use_case_studies}.

\section{Prior Frameworks and the Boundary They Assume}
\label{sec:prior}

Three bodies of work define what a dApp is today; each is examined for the boundary it places between the DU and the application, because that boundary decides which use cases it can express at all.

\textbf{Northeastern dApp framework.} Lacava~\etal~\cite{lacava2025dapps} pair an E3 agent embedded in OpenAirInterface with dApps that run as separate processes. Management uses E3AP, an ASN.1 service model compiled with asn1c in the E2 tradition, carried over ZeroMQ or TCP by default; the 2026 revision publishes the codec as a vendor-neutral library with SCTP among its link layers~\cite{libe3}. Data moves on a second plane: the agent publishes indications through ZeroMQ or a shared-memory reference, Python dApps subscribe, usually keeping only the newest message, and control actions return the same way to be applied at a safe point. The design is explicitly nonblocking; spectrum sharing, interference detection, and ISAC inference run on it~\cite{listen_while_talking,santhi2025interforan,isac_6gr}, and jBPF has been evaluated as a verified in-process hook for small functions~\cite{jbpf2024}. Boeira~\etal~\cite{boeira2026perf} characterize the same framework's performance: co-located containers add under 0.5\,ms to a control loop and separated containers 1 to 2\,ms, the CPU rather than the GPU saturates first, and about one core per dApp must be provisioned to keep a 10\,ms loop. Those are loop-level figures over a Python data plane and an RF simulator; the present study measures the carriers and codecs beneath such a loop, at microsecond resolution, with a cell on the air.

\textbf{NVIDIA Aerial dApps.} Villa~\etal~\cite{villa2026aerialdapp} add a dApp path to the CUDA-accelerated Aerial L1: a GPU-side export copies a slot's IQ or channel-estimate tensor out of the PHY's buffers to a second context, where an inference model runs and reports through a host process, within a budget of a few milliseconds. Its defining property is an export that leaves the PHY's own stream untouched.

\textbf{Vendor accelerators.} Commercial L1 stacks expose configurable pipelines and, increasingly, neural components~\cite{nvidia_aerial,cohenarazi2025ai_aerial}, but in-pipeline replacement of a stage by third-party code is not a supported contract in any of them; the standard-compliant neural receiver of~\cite{neural_rx_2024} ships as a vendor feature.

The first two groups share two properties: the dApp lives outside the L1 process, and the L1 owns the clock, so the application sees a copy and replies when it can. These make the frameworks portable, restartable, and safe to write in Python, and they fix the set of use cases the frameworks can express: a tap can observe a stage but cannot replace one, so a neural equalizer cannot sit on the far side of an export, and a scheduler decision cannot be applied ``at a safe point later'' when the slot is now. \ocudu{} keeps the observer boundary, implements E3AP over SCTP with the same asn1c toolchain, and adds two positions on the coupling spectrum so that the rows the boundary cannot express have a home.

\section{Two Axes, Three Classes, One Spectrum}
\label{sec:axes}

\subsection{The Axes That Change the Interface}
For every use case in the corpus of Section~\ref{sec:corpus}, two properties decide the interface before any model is chosen. The \emph{blocking contract} states who waits: in an \emph{inline} contract the PHY cannot progress until the function returns; in a \emph{bounded} contract the MAC or PHY control loop waits but has a conventional decision it can take at a deadline; in a \emph{nonblocking} contract nobody waits and the result is consumed later or dropped. The \emph{payload residency} states where the bytes are: grids, channel estimates, and LLRs are produced in GPU memory by the accelerated PHY~\cite{pennybacker2026cudaocudu} and stay there until the transport block is decoded; scheduler inputs are compact host structures the MAC already holds; spectrum windows and SRS responses are large and GPU-resident but read a few times per second.

The three classes of~\cite{pennybacker2026dappclasses} are the three non-empty combinations. \ClassA{}, \emph{Resident Inline L1}, is inline over resident tensors; \ClassB{}, \emph{Bounded Real-Time Control}, is bounded over compact host state; \ClassC{}, \emph{Asynchronous Observation and Advisory}, is nonblocking over either. The missing combinations are the ones physics excludes: an inline contract over exported data pays a copy and a wake on the slot's critical path, and a nonblocking contract over resident tensors is \ClassC{} with a lease.

\subsection{The Coupling Spectrum}
\begin{figure*}[t]
\centering
\begin{tikzpicture}[
  >=Latex, font=\sffamily\footnotesize,
  pos/.style={draw=#1, thick, rounded corners=2pt, fill=#1!8, align=center,
              text width=2.55cm, minimum height=1.55cm, inner sep=3pt},
  lab/.style={font=\sffamily\scriptsize, text=black!65, align=center, text width=2.55cm},
  end/.style={font=\sffamily\small\bfseries, align=center, text width=3.2cm},
]
\draw[->, very thick, black!60] (-0.6,0) -- (16.4,0);
\node[end, anchor=south west] at (-0.6,0.15) {ABI-standard end\\{\mdseries\scriptsize tight coupling, no copy, one process}};
\node[end, anchor=south east] at (16.4,0.15) {Protocol-standard end\\{\mdseries\scriptsize loose coupling, wire format, any process}};

\node[pos=clsA] (a) at (1.6,-1.5) {\textbf{\ClassA{}}\\Resident Inline L1\\[1pt]{\scriptsize direct C ABI on the host CUDA stream}};
\node[pos=clsB] (b) at (4.6,-1.5) {\textbf{\ClassB{}}\\Bounded RT Control\\[1pt]{\scriptsize direct call, validated intents, admitted deadline}};
\node[pos=clsC] (c1) at (7.6,-1.5) {\textbf{\ClassC{} in-process}\\[1pt]{\scriptsize leased views, latest-context cache}};
\node[pos=clsC] (c2) at (10.6,-1.5) {\textbf{\ClassC{} supervised}\\[1pt]{\scriptsize shm ring + eventfd, SEQPACKET control, CUDA-IPC pool}};
\node[pos=clsC] (c3) at (13.6,-1.5) {\textbf{\ClassC{} portable}\\[1pt]{\scriptsize E3AP/APER over SCTP, FlatBuffers profile, container or remote host}};

\node[lab, below=2pt of a] {pay: trust, ABI freeze\\get: 0 copies, slot deadline};
\node[lab, below=2pt of b] {pay: validation, timeout\\get: 100\,\us{} admitted loop};
\node[lab, below=2pt of c1] {pay: quiescence on unload\\get: no export, no wake};
\node[lab, below=2pt of c2] {pay: 1 copy, 1 wake\\get: crash isolation, restart};
\node[lab, below=2pt of c3] {pay: codec, kernel, network\\get: vendor neutrality, E2 tooling};

\foreach \x in {1.6,4.6,7.6,10.6,13.6} { \draw[black!60, thick] (\x,0.1) -- (\x,-0.1); }
\draw[black!55, thick] (9.2,-3.55) -- (9.2,-3.75) -- (15.0,-3.75) -- (15.0,-3.55);
\node[font=\sffamily\scriptsize, text=black!65, below=1pt] at (12.1,-3.75) {boundary used by prior dApp frameworks (observer with a reply channel)};
\end{tikzpicture}
\caption{The coupling spectrum. The three classes are positions on one axis, and \ClassC{} spans three of them. Prior frameworks occupy the right-hand half; the left-hand positions are what make inline and current-slot use cases expressible. Each position states what the dApp author pays and what the placement buys.}
\label{fig:spectrum}
\end{figure*}
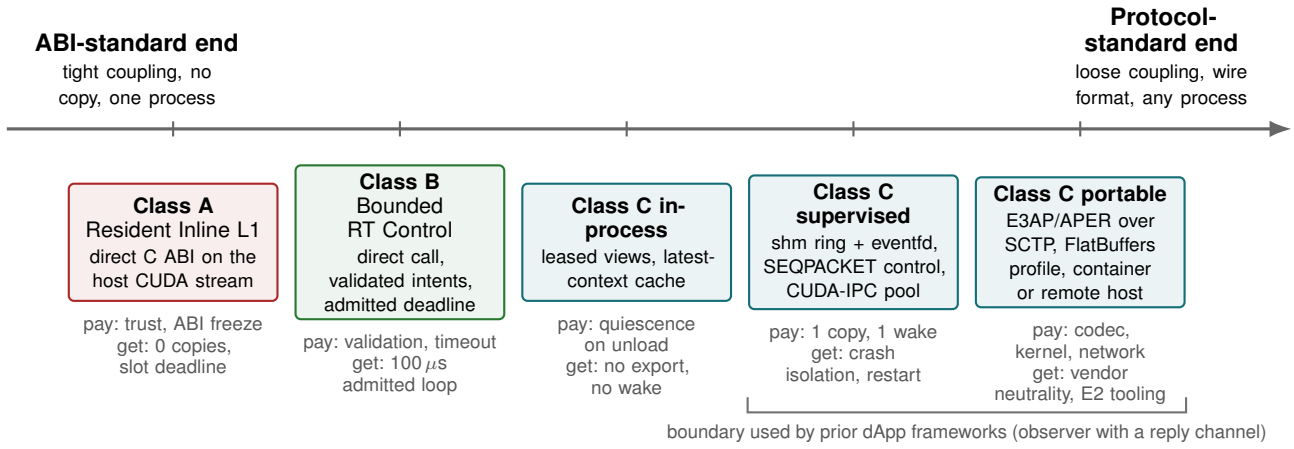

Fig.~\ref{fig:spectrum} presents the classes not as three boxes but as positions on one axis, the degree of coupling between the dApp and the DU, because that axis explains both the mechanism choices and the reason the platform keeps more than one.

At the \textbf{ABI-standard end} the dApp is a shared object loaded into the DU process and bound to a frozen, size-tagged C interface; the DU passes device pointers, a CUDA stream, and a deadline, and the dApp writes into DU-owned buffers and returns, with nothing copied or serialized. The cost is trust: the module runs with the DU's privileges and the ABI must not change under it. \ClassA{}, \ClassB{}, and the in-process \ClassC{} path live here. At the \textbf{protocol-standard end} the dApp is a process, a container, or a remote host that speaks a wire protocol. E3AP over SCTP is that protocol: the carrier the O-RAN E2 interface standardized~\cite{oran_e2gap,rfc4960}, one the Northeastern codec library supports, and the one several large vendors prefer for RAN control planes because their tooling, security posture, and conformance practice already exist for it. The cost is a codec and a kernel round trip per message; what it buys is vendor neutrality, language independence, crash isolation, and placement wherever the operator's security model requires. Between the ends sits the \textbf{supervised} position: a separate process on the same host sharing memory with the DU, paying one copy and one wakeup for crash isolation and restart.

The rest of the paper uses a small vocabulary for the positions. A \emph{native} dApp is a module in the DU process; for \ClassC{} it reads a \emph{leased} view, a zero-copy reference the DU refuses to invalidate until the lease returns. A \emph{supervised} dApp is a worker process on the same host fed through a shared-memory ring and restarted by the DU when it dies. A \emph{portable} dApp is any E3 peer. The \emph{latest-context cache} is a host-owned structure into which a \ClassC{} dApp publishes and from which a \ClassB{} call reads, so that an observation reaches a decision without a message.

Three consequences follow. The classes are not a ranking: a position is right or wrong for a use case. \ClassC{} is not ``observer only'': its contract permits every position, and it closes loops through the latest-context cache and through E3 control actions. And \emph{optionality} is itself a requirement: a group already running Python dApps over E3 should be able to point them at \ocudu{} unchanged, a vendor shipping a neural receiver needs the left end, and an operator with a hardened-container policy needs the right end for anything it did not write. Section~\ref{sec:meas} prices each position so the choice can be made per use case, and Section~\ref{sec:meas:b} prices the one combination the spectrum admits but the platform does not ship, a supervised \ClassB{}.

\section{The Use-Case Corpus, Audited}
\label{sec:corpus}

\begin{figure}[t]
\centering
\includegraphics[width=\columnwidth]{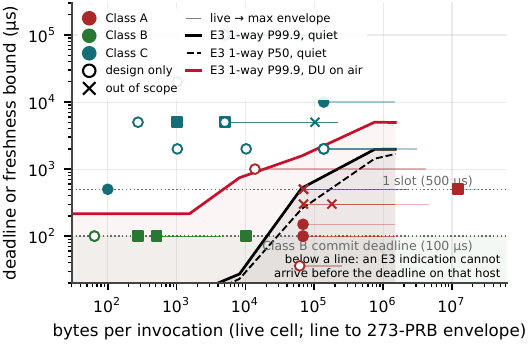}
\caption{The 39 runtime rows by bytes per invocation and deadline. Filled markers are shipped or live, hollow are design only, crosses are out of scope. The lines are the measured one-way delivery of an E3AP indication (APER over SCTP) on the quiet host, P99.9 solid black and P50 dashed, and with the DU on the air, P99.9 red; a row below a line cannot be served across that boundary within its deadline on that host. The \ClassB{} rows sit between the two lines, which is the whole \ClassB{} argument: the boundary fits on an idle host and not on the host the DU runs on.}
\label{fig:corpus_map}
\end{figure}

The corpus of~\cite{pennybacker2026dappclasses} listed 45 use cases drawn from the O-RAN nGRG report on dApps~\cite{oran_ngrg}, the dApp literature~\cite{lacava2025dapps,listen_while_talking,santhi2025interforan,villa2026aerialdapp}, and AI-native PHY work. Every entry was re-examined against the 5G NR timing that bounds it (a 500\,\us{} slot at 30\,kHz, retransmission DCI after $k_2 \ge N_2$, SRS periods of 5--40\,ms, beam-failure timers of 10--200\,ms), the bytes it moves at the released testbed shape (51\,PRB, two receive ports) and at the 273-PRB, four-port envelope, and the contract the released runtime offers. Rows mixing a per-slot half with a policy half were split, rows whose input the DU never sees or whose actuator sits inside the RU were moved out, and ten rows the platform made possible, such as the observation-to-avoidance composition, were added. The audited corpus has 39 runtime rows (13 \ClassA{}, 9 \ClassB{}, 17 \ClassC{}), 9 xApp/rApp rows acting through typed configuration or RAN control, one research row that needs a multi-DU substrate, and five compositions that flow only through host-owned caches and authorities. Of the runtime rows, 14 are shipped or demonstrated live, 12 are design only, and 13 are out of scope for the released platform. The Appendix lists every row; the formulas and sources behind each are in the study repository~\cite{ocudu_use_case_studies}. Fig.~\ref{fig:corpus_map} places every row against the two axes and the measured cost of the observer boundary.

A verdict needs a definition of feasible. A row is \emph{feasible} across a boundary when the measured P99.9 of the mechanism that boundary uses, at the row's byte count and with the DU on the air, is inside the row's deadline; \emph{infeasible} when a path exists and that tail is outside it; and \emph{inexpressible} when no path returns the result into the consuming stage at all. The DU-on-air condition is the one that counts because a bounded contract is defined by its tail under the load the DU itself creates (Section~\ref{sec:meas:b}).

Three facts carry the argument, and the verdicts behind them use the byte and deadline arithmetic of the corpus tables and the measurements of Section~\ref{sec:meas}. Eleven of the 13 \ClassA{} rows are inexpressible in either external framework, for a reason unrelated to speed: an indication carries data outward, and nothing brings a channel estimate, an equalized tensor, or an LLR tensor back into the PUSCH chain of the same slot; the remaining two, PRACH detection and SRS estimation, run rarely enough or on tensors small enough that they do not discriminate. Every \ClassB{} row has a return path in the observer-only framework, the E3 control action, and none in the export framework, which documents no decision-boundary hook; on the quiet host that return path fits the 100\,\us{} budget, and with the DU on the air its tail does not (Section~\ref{sec:meas:b}), which is why Fig.~\ref{fig:corpus_map} draws the boundary under both conditions. Every \ClassC{} row is feasible in all three frameworks, which is the case for keeping the protocol-standard end: for 17 of 39 rows an out-of-process E3 client is a correct implementation, and removing it would trade portability for nothing.

Table~\ref{tab:drivers} lists the rows that most sharply separate the frameworks, with the mechanism each would have to use and its measured or published cost. Its \ClassC{} rows show where the platform's contribution lies once expressibility is settled: producer-side cost, survival across consumer restarts, and the two return paths by which an observation re-enters control (the interference map consumed by the scheduler and the quiet-period reservation admitted by a host authority) without a message on the decision path.

\begin{table*}[t]
\centering
\caption{Discriminating rows of the audited corpus. Costs are P50 unless stated; \emph{meas.} rows come from the suite in Section~\ref{sec:meas} or from the released runtime, \emph{publ.} from the cited papers. ``Inexpressible'' means no return path into the consuming stage exists.}
\label{tab:drivers}
\scriptsize
\setlength{\tabcolsep}{3pt}
\begin{tabularx}{\textwidth}{@{}L{2.9cm}L{2.5cm}YYY@{}}
\toprule
\textbf{Row} & \textbf{Bytes, budget} & \textbf{Observer-only external (E3AP/SCTP + msg.\ data plane)} & \textbf{External GPU pipeline (D2H or D2D export)} & \textbf{\ocudu{} resident or bounded (in-process ABI)} \\
\midrule
\textcolor{clsA}{A-03} neural receiver to LLRs & 1.47\,MB in, up to 4.4\,MB out per slot; $\le$500\,\us{} occupancy & Inexpressible. For scale, a 734\,KB indication costs \DrvEInd{} one way with the real APER codec over SCTP (meas.) & Inexpressible. Export of 1.47\,MB to host costs \DrvDtoH{} before inference (meas.) & Feasible at the live shape: \ChkARxLive{} P50 / P99.9 (meas.), 92.5 / 105.1\,\us{} runtime checkpoint; 273-PRB kernels not yet qualified \\
\textcolor{clsA}{A-02} neural equalizer & 68.5\,KB in (live); same-invocation consumption & Inexpressible & Inexpressible & Feasible: 45--52\,\us{} P50; 260{,}000+ invocations on air, 0 fallbacks \\
\midrule
\textcolor{clsB}{B-01} scheduler intents & $\le$4.2\,KB in, $\le$2.3\,KB out; 100\,\us{} to commit & Infeasible under DU load: SCTP loopback round trip \DrvSctpRT{} P99.9 quiet, \ContBSctp{} with the DU on the air (meas.); about 400\,\us{} in published loops & Inexpressible: no decision-boundary hook & Feasible: direct call \DrvAbiRT{} P99.9 quiet, \ContBAbi{} with the DU on the air, 20{,}000 validated calls (meas.); 3.6\,\us{} on the runtime \\
\textcolor{clsB}{B-04} avoidance from the C-01 map & $\le$275\,B context; 100\,\us{} to commit & Infeasible: the map is an indication, its consumption is the B-01 path & Inexpressible & Feasible: latest-context cache, 3 bounded acquire attempts; demonstrated live \\
\midrule
\textcolor{clsC}{C-01} spectrum sensing & 137\,KB (live) to 1.47\,MB per slot; no deadline & Feasible: 1.5\,KB decimated indications in published loops; a full grid costs \DrvEGrid{} one way (meas.) & Feasible: about 150\,\us{} C API, 350\,\us{} gRPC at 717\,KB per slot (publ.) & Feasible on all 3 positions; producer critical section \DrvRingProd{} P99.9, independent of the subscriber (meas.) \\
\textcolor{clsC}{C-02} quiet reservation & about 100\,B intent; $\ge$1 slot ahead & Feasible as an E3 control message; admission and duty bound absent from the published procedure set & Inexpressible: no scheduler authority & Feasible: host reservation authority with clamp and duty budget; shipped \\
\textcolor{clsC}{C-03} SRS-ISAC & $\le$210\,KB per SRS occasion; $\le$ SRS period & Feasible: \DrvESrs{} one way at 64\,KB (meas.) & Feasible: about 2.3\,ms cuSense (publ.) & Feasible: CUDA-IPC pool, 1 device-to-device snapshot; demonstrated live \\
\bottomrule
\end{tabularx}
\end{table*}

\section{Measured Interface Costs}
\label{sec:meas}

\subsection{Method}
\label{sec:meas:method}
All rows were measured on one NVIDIA DGX Spark (GB10, 20-core Arm host, coherent CPU--GPU memory) under two conditions: \emph{quiet}, with the host otherwise idle, and \emph{contention}, with the released \ocudu{} cell on the air on the same host and sharing its GPU. The text and figures report the quiet run; Table~\ref{tab:conditions} sets the two side by side for the rows the argument rests on. Benchmarks ran pinned to the six cores the DU does not use, with memory locked, a 10\,\% warmup, and raw nearest-rank percentiles with no tail trimming; \code{CLOCK\_MONOTONIC} above 1\,\us{} and the cycle counter below it. Every CSV row carries its contract, placement, copies and system calls on the measured path, and an evidence kind, \emph{measured} or \emph{published baseline}. Where earlier drafts of this study modeled a cost, for example E3 as a fixed header plus a per-byte slope, the real mechanism is now measured, and rows once labeled inter-process that ran inside one process now fork.

The mechanisms are the ones the frameworks ship. The E3 rows link the \ocudu{} asn1c APER codec and SCTP transport from the SDK, with the message families and 64\,KiB fragmentation of the E3AP wire contract, and the Northeastern libe3 codec beside it. The \ocudu{} inter-process rows use the released ring layout (128\,B header, 64\,B slot headers, 64\,B-aligned stride, try-lock guard, eventfd signal) and its SEQPACKET control channel with credential checks on every message. The Northeastern-style rows use ZeroMQ publish/subscribe over \code{ipc://} across a fork with the high-water mark and conflating-subscriber options that framework's Python dApps use; the NVIDIA-style rows use a device-to-host copy into pinned memory, a shared-memory hand-off, and a ZeroMQ notification. Payloads are the corpus's: 34 and 68\,KB for a 51-PRB slot as 16-bit and 32-bit complex, 734\,KB and 1.47\,MB for the 273-PRB, four-port envelope, 64\,KB for an SRS occasion, and a 16-candidate scheduler request of 3.2\,KB returning 1.2\,KB of intents.

\begin{table}[t]
\centering
\caption{The same rows under the two host conditions (\us{}). Quiet: host otherwise idle. Contention: the released cell on the air on the same host, sharing the GPU, with the benchmark pinned to the cores the DU does not use.}
\label{tab:conditions}
\scriptsize
\setlength{\tabcolsep}{3pt}
\begin{tabular}{@{}lrrrr@{}}
\toprule
& \multicolumn{2}{c}{\textbf{Quiet}} & \multicolumn{2}{c}{\textbf{Contention}} \\
\cmidrule(lr){2-3}\cmidrule(lr){4-5}
\textbf{Row} & \textbf{P50} & \textbf{P99.9} & \textbf{P50} & \textbf{P99.9} \\
\midrule
\textcolor{clsA}{A} GPU-resident inline, 1.47\,MB grid (floor) & 58 & 119 & 13 & 801 \\
\textcolor{clsB}{B} direct ABI, validated call & 0.27 & 0.29 & 0.27 & 5.22 \\
\textcolor{clsB}{B} same request over SCTP loopback & 13 & 16 & 13 & 181 \\
\textcolor{clsC}{C} ring producer critical section, 68\,KB & 1.97 & 2.34 & 1.97 & 167 \\
\textcolor{clsC}{C} ring consumer wake, 68\,KB & 3.92 & 99 & 4.91 & 1.62\,ms \\
\textcolor{clsC}{C} SEQPACKET control RTT/2, 256\,B & 3.20 & 4.26 & 3.25 & 69 \\
\textcolor{clsC}{C} E3AP indication 64\,KB one way & 255 & 520 & 207 & 1.57\,ms \\
\textcolor{clsC}{C} E3AP control round trip & 11 & 13 & 12 & 492 \\
\bottomrule
\end{tabular}
\end{table}

\subsection{The Protocol-Standard End: APER and SCTP}
\label{sec:meas:e3}
\begin{figure}[t]
\centering
\includegraphics[width=\columnwidth]{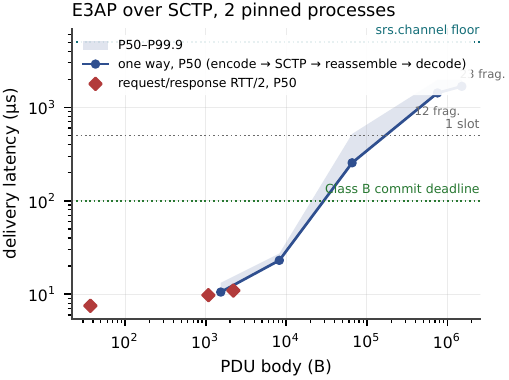}
\caption{The protocol-standard end, measured with the real codecs and carrier. One-way delivery of E3AP indications (asn1c APER encode, SCTP send, reassembly, decode) over an association between two pinned processes, and the control-plane request/response exchanges, against the deadlines that matter for each class. Codec costs alone are given in the text.}
\label{fig:e3}
\end{figure}

Fig.~\ref{fig:e3} gives the cost of the protocol-standard end with nothing modeled. The codec rows encode and decode real E3AP PDUs with the \ocudu{} asn1c APER codec: setup, subscription, indications with bodies from 64\,B to a fragmented 734\,KB grid, a 2.2\,KB scheduler control action, and an acknowledgment. Two regimes appear. For control-sized messages the cost is the envelope, \EThreeCtlEnc{} to encode and \EThreeCtlDec{} to decode a control action, independent of the few hundred bytes of body, and the Northeastern codec tracks it closely. For indications the cost grows linearly at \EThreeSlope{} per byte because APER copies an \code{OCTET STRING} body bit-aligned: a 64\,KB SRS indication costs \EThreeSrsEnc{} and a 734\,KB grid \EThreeGridEnc{} to encode, with decode near \EThreeDecRatio{} the encode cost. The FlatBuffers data profile on the second association costs \EThreeFbGrid{} for the same grid, which is why the platform has two associations.

The carrier rows send those PDUs between two pinned processes over an SCTP one-to-one association on loopback, one PDU per message, fragmented at 64\,KiB as the E3AP wire contract requires, and time one-way delivery through reassembly and decode. A 1.5\,KB indication arrives in \EThreeOneWaySmall{} at the median; the 734\,KB grid, 12 fragments, in \EThreeOneWayGrid{} (\EThreeOneWayGridTail{} at P99.9); the 1.47\,MB four-port grid in \EThreeOneWayBig{}. A control round trip on stream 0 with the real codec at both ends is \EThreeRTFifty{} at the median and \EThreeRT{} at P99.9; the published loops of about 400\,\us{}~\cite{lacava2025dapps} sit between these compiled figures and the same path through a Python decoder. With the DU on the air the medians barely move and the tails do: the 1.5\,KB indication reaches \ContEOneWaySmall{} at P99.9, the 64\,KB SRS indication \ContEOneWaySrs{}, and the control round trip \ContERT{} (Table~\ref{tab:conditions}).

These numbers draw the lines on the corpus map of Fig.~\ref{fig:corpus_map}. Every \ClassC{} row lies above both, most by an order of magnitude, so an E3 client is a correct implementation of all 17. Every \ClassA{} row lies below the quiet-host line, so even before the missing return path is counted the export alone does not fit. Every \ClassB{} row lies above the quiet-host line and below the DU-on-air line: the boundary fits the 100\,\us{} budget on an idle machine and not on the machine the DU runs on. That is the whole case for the right-hand end: it costs nothing the asynchronous rows cannot afford, and no codec optimization moves the lines, because they are set by the copy, the wake, and the association.

\subsection{\ClassC{} Egress: The Producer Rule}
\label{sec:meas:c}
\begin{figure*}[t]
\centering
\includegraphics[width=0.92\textwidth]{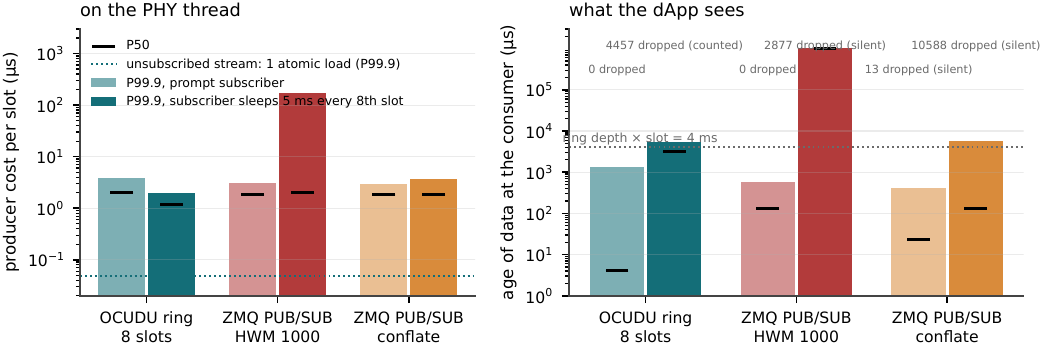}
\caption{Producer-side cost per 68\,KB slot publication, P50 (tick) and P99.9 (bar), with a prompt subscriber and with a subscriber that sleeps 5\,ms every 8th slot. The \ocudu{} ring's producer cost does not move; a full ring is a counted drop. ZeroMQ's producer inherits the subscriber's backlog.}
\label{fig:producer}
\end{figure*}

A \ClassC{} publisher runs on the PHY thread. Whatever the consumer does, the producer's cost per slot must stay bounded and small, and the data the consumer eventually sees must have a known age. Fig.~\ref{fig:producer} tests both properties for the \ocudu{} positions and for the ZeroMQ publish/subscribe path of the observer frameworks, at the live 68\,KB slot.

The supervised ring is the released layout in a two-process harness: try-lock guard, reclaim read, memcpy into the next 64\,B-aligned slot, release store, eventfd write, \RingSyscalls{} system calls per publication. Its critical section costs \RingProdFifty{} at the median for 68\,KB and \RingProdBig{} for the 1.47\,MB envelope, which is the memcpy; when the ring is full the producer returns in \RingProdFull{} with a counted drop and never waits. In the slot-timed run the producer publishes every 500\,\us{} for \SlotN{} slots while the consumer either drains promptly or sleeps 5\,ms on every 8th slot. With the sleeping consumer the ring drops \SlotDrops{} slots across \SlotStalls{} stalls, about two per stall, which is exactly the excess of a 5\,ms stall over an 8-slot ring, and the age of the newest data the consumer sees is bounded at \SlotRingAge{} median and \SlotRingAgeMax{} P99.9 by the ring depth. An unsubscribed stream costs the producer one relaxed atomic load, \SlotIdle{}.

ZeroMQ over \code{ipc://} publishes the same 68\,KB in \ZmqProdFifty{} at the median, the same producer cost; the difference is what happens to the data. With the default high-water mark the socket queues everything a slow subscriber has not read, so the age of the data it eventually sees grows to \ZmqAgeSlow{} at the median and \ZmqAgeSlowMax{} at P99.9, with no drop reported; with the conflating subscriber option the Python dApps use, the age stays at one slot but \ConflateDrops{} slots are discarded inside the socket with no count anywhere. The ring is not faster than the socket; it is \emph{accountable}: staleness is bounded by a configured depth and every drop is counted, which is what an operator needs to know whether an advisory loop is still advising. On the quiet host the ring producer's P99.9 is \RingProd{}; with the DU on the air it is \ContRingProd{}, and the consumer's wake tail grows from \RingWake{} to \ContRingWake{} (Table~\ref{tab:conditions}), because the bench cores then also carry DU worker threads and enter deep idle states between slots.

The SEQPACKET control channel, credentials verified on every receive, completes a 256\,B request/response in \SeqRT{} at the median and \SeqRTTail{} at P99.9 and a heartbeat in \SeqHb{}, against \SeqZmq{} for ZeroMQ request/reply: the cost of noticing a dead worker within one heartbeat and restarting it without touching the ring. The consumer's wake from eventfd to a leased slot is \RingWakeFifty{} at the median and \RingWake{} at P99.9. The native lease costs the producer only the unsubscribed check.

\subsection{\ClassB{}: One Request, Six Boundaries}
\label{sec:meas:b}
The \ClassB{} question is whether a process boundary can sit inside a 100\,\us{} decision. All six paths carry the same request, a 16-candidate scheduler input of 3{,}240\,B (the released \code{scheduler\_input\_v1} with two 275-bit masks) returning 16 intents of 72\,B. Each path invokes the same dApp function through the released interface's \code{invoke} pointer: a direct call; a heap SPSC pair between two pinned threads; a shared-memory SPSC across a fork; ZeroMQ request/reply over \code{ipc://} and again over TCP loopback; and an SCTP one-to-one association on loopback with real message boundaries, the carrier an E3 control action would use. The direct call is timed with the DU's admission checks included, which cover structure size and ABI major, candidate identity, allow flags, PRB range, MCS bound, and deadline. Each path runs 20{,}000 requests on separate pinned cores under three conditions: the quiet host, the quiet host with a 50\,\us{}-on, 50\,\us{}-off busy thread on the responder's core, and the host with the DU on the air.

\begin{figure}[t]
\centering
\includegraphics[width=\columnwidth]{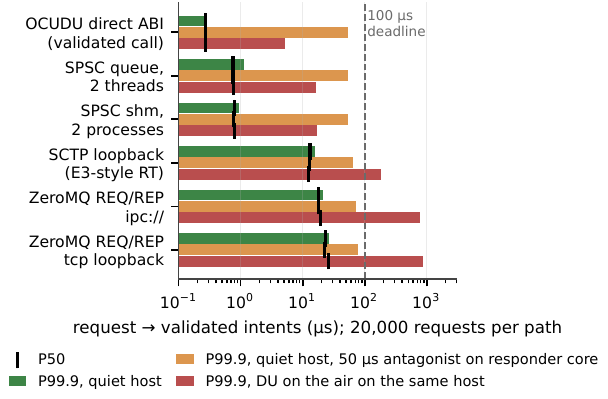}
\caption{One \ClassB{} request (3.2\,KB in, 1.2\,KB of intents out) over six boundaries, 20{,}000 requests each, under three host conditions. Bars are P99.9, ticks are P50; the dashed line is the admitted deadline. On the quiet host every path fits; with the DU on the air only the paths without a wake keep their tails.}
\label{fig:classb}
\end{figure}

Fig.~\ref{fig:classb} shows the result. On the quiet host every carrier meets the deadline: the direct call completes in \MechBAbiFifty{} at the median and \MechBAbi{} at P99.9 with validation included, the shared-memory SPSC in \MechBShmFifty{} and \MechBShm{}, SCTP in \MechBSctpFifty{} and \MechBSctp{}, ZeroMQ over \code{ipc://} in \MechBZmqFifty{} and \MechBZmq{}. The antagonist adds about 50\,\us{} to every tail (\MechBSctpAnt{} for SCTP, \MechBZmqAnt{} for ZeroMQ, \MechBAbiAnt{} for the direct call, whose caller shared the contested core); all still fit. With the DU on the air the picture changes: the direct call moves to \ContBAbi{} at P99.9 and the SPSC to \ContBShm{}, while the message paths move to \ContBSctp{} (SCTP) and \ContBZmq{} (ZeroMQ) and miss the deadline on \ContBSctpMiss{} and \ContBZmqMiss{} of requests with the responder core otherwise idle. APER encode and decode at both ends of the SCTP path add up to \MechBE{} at the median on the quiet host, a floor for a compiled E3 control loop; the published loops of about 400\,\us{}~\cite{lacava2025dapps} include a Python decoder and the agent's data plane, and report no tails.

The reading is not that ZeroMQ or SCTP are slow; on an idle host they would do. It is that a bounded contract is defined by its tail under the load the DU itself creates, and the tail of any path with a wake in it is then set by the kernel and by whatever else the DU runs. Nor is a miss free because the contract has a fallback. The conventional decision is taken at the deadline, but the MAC has already spent the wait, and a \ContBZmq{} tail is longer than the slot: the request outlives the decision it was for, the reply must be discarded on arrival, and the intents for the next slot are computed from a context one slot stale. A miss rate of \ContBSctpMiss{} at three UEs on a 51-PRB cell is the floor of that cost, not its ceiling, since Boeira~\etal~\cite{boeira2026perf} show the same framework's CPU saturating before its GPU as dApps are added.

The busy-polling shared-memory SPSC deserves its own reading, because it is a process boundary that stays inside the budget under load, \ContBShm{} at P99.9 with \ContBShmMiss{} misses. It is the supervised \ClassB{} position the spectrum admits: one copy of 3\,KB, one copy of 1\,KB back, crash isolation, and no wake, at the price of a core the responder must spin on and the DU therefore loses. The platform defines that position as an alternative and does not ship it, for two reasons the measurements make concrete. The validation a \ClassB{} call needs is the same on either side of the boundary, 72-byte intents range-checked field by field, so isolation buys protection against a crash but not against a wrong answer, and the circuit breaker already bounds the crash. And the direct call is still an order of magnitude tighter, \ContBAbi{} against \ContBShm{}, on a budget that the feature block and commit also have to share. The one path whose tail stays inside the budget on the host that matters without reserving a core is the direct call; the price, validating 72\,B intents, is small (Section~\ref{sec:design}).

\subsection{jBPF: A Verified Hook, and Why It Was Not Adopted}
\label{sec:meas:jbpf}
jBPF~\cite{jbpf2024} runs verifier-checked codelets inside the DU at named hooks, the one prior mechanism that is not an observer with a reply channel, and the suite measures it as shipped. The hook is cheap, \JbpfHookSmall{} for a 64\,B context and \JbpfHookMid{} for 16\,KB, rising to \JbpfHookGrid{} for a 1.47\,MB grid the codelet must read in full. The limit is what the codelet can do with its result: its output goes into jBPF's own ring, which an I/O thread drains for a consumer, so a request/response through the hook costs \JbpfRT{} at the median, set by that thread's polling interval, and in every published use the consumer is an external process, so the tap adds a copy into the ring before the copy across the boundary. The verifier also excludes device pointers and CUDA calls, which \ClassA{} needs, and bounds the loops a scheduler policy would write. The platform therefore keeps jBPF as an evaluated option for tiny \ClassB{} taps and adopts, for the same contract, a direct call whose output lands in the caller's structure and is validated field by field, with no ring, no I/O thread, and no second copy.

\subsection{\ClassA{}: What an Export Costs Before the Model Runs}
\label{sec:meas:a}
The \ClassA{} rows are inexpressible on an external boundary for the structural reason of Section~\ref{sec:corpus}, so the measurement answers a narrower question: were a framework extended with a return path, what would the export alone cost before any inference ran? On the coherent-memory host a device-to-host copy of a 1\,MB grid into pinned memory takes \MechADtoHFifty{} at the median and \MechADtoH{} at P99. The full NVIDIA-style path (copy, shared-memory stage, ZeroMQ notification) takes \MechANvFifty{} at the median and \MechANvNN{} at P99 for the 1.47\,MB four-port grid, against a 300\,\us{} planning budget. Both paths measure the outbound leg alone. A stage needs the result back, so the round trip is at least twice the outbound figure, about 450\,\us{} at the median for the NVIDIA-style path before a single model FLOP, and no framework provides the return copy. The in-process figure is the cost of enqueueing the dApp's kernels on the lane's stream and recording the completion event, measured with the released packages on the exclusive GPU. At the live shape (51\,PRB, two layers, 256-QAM), the channel estimator submits in \ChkAStreamFifty{} at the median and the reference receiver completes in \ChkARxLive{} P50 / P99.9, inside its 150\,\us{} qualification deadline. At the 273-PRB envelope the same receiver submits in \ChkARxFullSubmit{} but completes in \ChkARxFull{}, so the reference kernels, not the interface, are what remains to be qualified at that shape. The framework-path suite shows the same thing at every payload from 256\,KB to 23\,MB: residency costs a launch, export costs a copy proportional to the tensor, and at the 64-port envelope the copy alone (\MechASixtyFour{} at 23.5\,MB) exceeds the slot.

\subsection{The Frameworks at Matched Contracts}
\label{sec:meas:paths}
Fig.~\ref{fig:paths} closes the section by putting the frameworks side by side under the harness above. For each of three workloads, one per class, the figure shows the median and P99.9 of each framework's path at the workload's byte count against its budget; published anchors from the cited papers appear as marks without percentiles. The picture is the one the corpus predicted. On the \ClassA{} workload only the resident path is inside the budget, and the other frameworks' paths are shown hatched because they are outside their authors' stated scope. On the \ClassB{} workload the resident path is more than two orders of magnitude inside the budget and the observer path is outside it. On the \ClassC{} workload all three are inside the budget by a wide margin, and the differences among them are the producer-side and restart properties measured above rather than feasibility.

\begin{figure*}[t]
\centering
\includegraphics[width=0.9\textwidth]{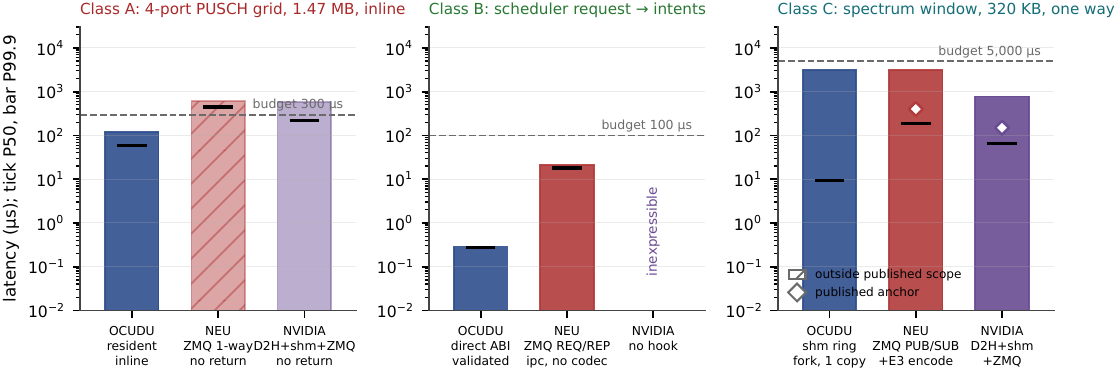}
\caption{Contract-matched framework paths, one workload per class, P50 (tick) and P99.9 (bar) against the workload budget (dashed). Hatched bars are outside the framework's published scope and are shown as sensitivity only; marks without bars are published anchors.}
\label{fig:paths}
\end{figure*}

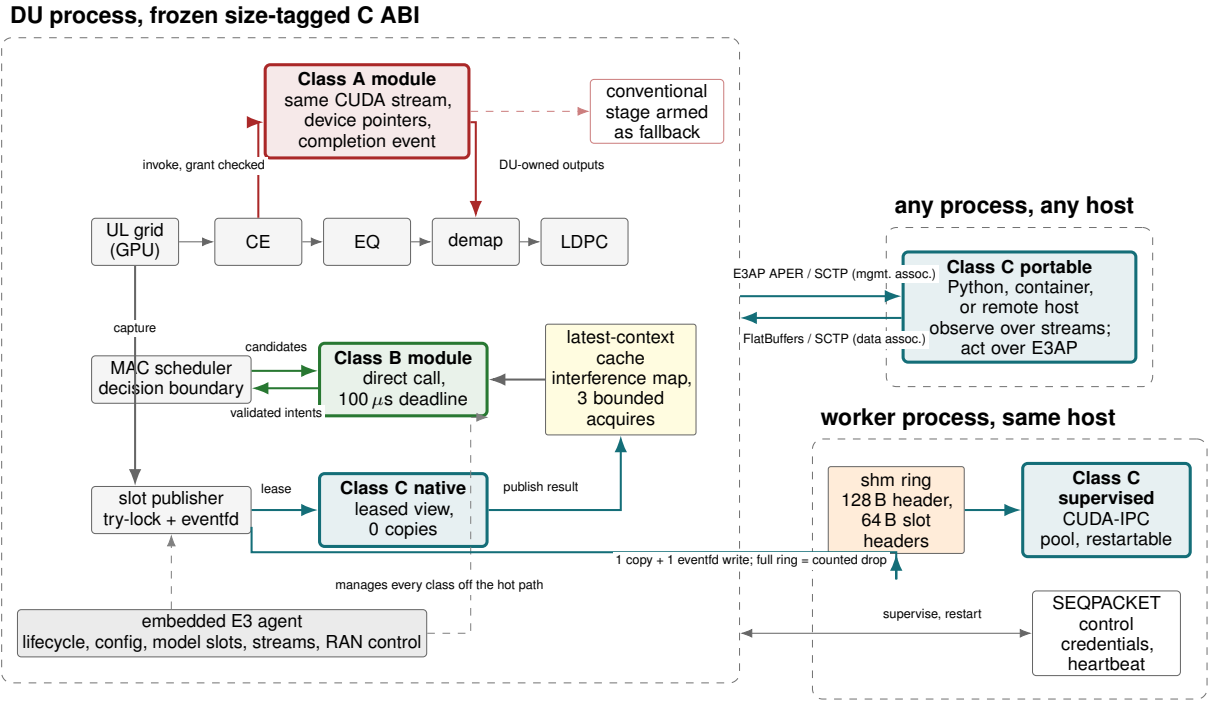
\begin{figure*}[t]
\centering
\resizebox{0.88\textwidth}{!}{%
\begin{tikzpicture}[x=1cm, y=1cm, >=Latex, font=\sffamily\scriptsize,
  stage/.style={draw=black!60, fill=black!4, rounded corners=1.5pt, minimum height=6.5mm, minimum width=12mm, align=center, inner sep=2pt},
  dapp/.style={draw=#1, very thick, fill=#1!10, rounded corners=2pt, minimum height=8mm, align=center, inner sep=3pt, text width=21mm},
  store/.style={draw=black!60, fill=yellow!15, rounded corners=1pt, align=center, inner sep=2.5pt, minimum height=6mm},
  proc/.style={draw=black!55, dashed, rounded corners=4pt, inner sep=6pt},
  flow/.style={->, thick, #1}, flow/.default=black!70,
  lbl/.style={font=\sffamily\tiny, fill=white, inner sep=1pt},
]
\node[stage] (grid) at (0.8,0) {UL grid\\(GPU)};
\node[stage] (ce) at (2.5,0) {CE};
\node[stage] (eq) at (4.0,0) {EQ};
\node[stage] (dem) at (5.5,0) {demap};
\node[stage] (ldpc) at (7.0,0) {LDPC};
\foreach \a/\b in {grid/ce, ce/eq, eq/dem, dem/ldpc} { \draw[->, black!60] (\a) -- (\b); }
\node[dapp=clsA, text width=26mm] (a) at (4.0,1.8) {\textbf{\ClassA{} module}\\same CUDA stream, device pointers, completion event};
\draw[flow=clsA] (ce.north) |- ($(a.west)+(0,-0.15)$);
\node[lbl] at (1.75,1.05) {invoke, grant checked};
\draw[flow=clsA] ($(a.east)+(0,-0.15)$) -| (dem.north);
\node[lbl] at (6.55,1.05) {DU-owned outputs};
\node[stage, draw=clsA!60, fill=white, text width=17mm] (fb) at (8.0,1.8) {conventional stage armed as fallback};
\draw[->, thin, clsA!60, dashed] (a.east) -- (fb.west);
\node[stage, minimum width=22mm] (mac) at (1.3,-1.9) {MAC scheduler\\decision boundary};
\node[dapp=clsB] (b) at (4.5,-1.9) {\textbf{\ClassB{} module}\\direct call, 100\,\us{} deadline};
\draw[flow=clsB] ($(mac.east)+(0,0.12)$) -- ($(b.west)+(0,0.12)$); \node[lbl] at (2.75,-1.45) {candidates};
\draw[flow=clsB] ($(b.west)+(0,-0.12)$) -- ($(mac.east)+(0,-0.12)$); \node[lbl] at (2.75,-2.35) {validated intents};
\node[store, text width=19mm] (ctx) at (7.5,-1.9) {latest-context cache\\interference map,\\3 bounded acquires};
\draw[flow=black!60] (ctx.west) -- (b.east);
\node[stage, minimum width=22mm] (pub) at (1.3,-3.7) {slot publisher\\try-lock + eventfd};
\draw[flow=black!60] (grid.south) -- (grid.south |- pub.north); \node[lbl] at (0.8,-1.2) {capture};
\node[dapp=clsC] (cn) at (4.5,-3.7) {\textbf{\ClassC{} native}\\leased view, 0 copies};
\draw[flow=clsC] (pub.east) -- (cn.west); \node[lbl] at (2.75,-3.4) {lease};
\draw[flow=clsC] (cn.east) -| (ctx.south); \node[lbl] at (6.4,-3.4) {publish result};
\node[stage, minimum width=32mm, fill=black!8] (agent) at (2.0,-5.4) {embedded E3 agent\\lifecycle, config, model slots, streams, RAN control};
\draw[->, thin, black!50, dashed] (agent.north -| pub.south) -- (pub.south);
\draw[->, thin, black!50, dashed] (agent.east) -- ++(0.6,0) |- (b.south east) ;
\node[lbl] at (5.0,-4.75) {manages every class off the hot path};
\begin{scope}[on background layer]
\node[proc, fit=(grid)(ldpc)(a)(fb)(mac)(b)(ctx)(pub)(cn)(agent), inner ysep=10pt] (du) {};
\end{scope}
\node[font=\sffamily\small\bfseries, anchor=south west] at (du.north west) {DU process, frozen size-tagged C ABI};
\node[store, fill=orange!15, text width=17mm] (ring) at (11.3,-3.7) {shm ring\\128\,B header, 64\,B slot headers};
\node[dapp=clsC] (cs) at (14.2,-3.7) {\textbf{\ClassC{} supervised}\\CUDA-IPC pool, restartable};
\node[stage, text width=19mm, fill=white] (seq) at (14.2,-5.4) {SEQPACKET control\\credentials, heartbeat};
\draw[flow=clsC] (ring.east) -- (cs.west);
\draw[flow=clsC] (pub.south east) ++(0,0.1) -- ++(0,-0.35) -| ($(ring.south)+(0,-0.35)$) -- (ring.south);
\node[lbl] at (9.3,-4.4) {1 copy + 1 eventfd write; full ring = counted drop};
\draw[<->, thin, black!60] (agent.east -| du.east) -- (seq.west); \node[lbl] at (11.8,-5.15) {supervise, restart};
\begin{scope}[on background layer]
\node[proc, fit=(ring)(cs)(seq), inner ysep=9pt] (wk) {};
\end{scope}
\node[font=\sffamily\small\bfseries, anchor=south west] at (wk.north west) {worker process, same host};
\node[dapp=clsC, text width=30mm] (cp) at (13.0,-0.9) {\textbf{\ClassC{} portable}\\Python, container, or remote host\\observe over streams; act over E3AP};
\draw[flow=clsC] (du.east |- cp.west) ++(0,0.15) -- ($(cp.west)+(0,0.15)$);
\draw[flow=clsC] ($(cp.west)+(0,-0.15)$) -- ($(du.east |- cp.west)+(0,-0.15)$);
\node[lbl] at (10.45,-0.45) {E3AP APER / SCTP (mgmt.\ assoc.)};
\node[lbl] at (10.45,-1.35) {FlatBuffers / SCTP (data assoc.)};
\begin{scope}[on background layer]
\node[proc, fit=(cp), inner ysep=9pt] (pf) {};
\end{scope}
\node[font=\sffamily\small\bfseries, anchor=south west] at (pf.north west) {any process, any host};
\end{tikzpicture}}
\caption{The three classes inside and around the \ocudu{} DU. \ClassA{} is invoked on the PUSCH lane's own stream and writes DU-owned outputs; \ClassB{} is a direct call at the scheduler's decision boundary returning validated intents; \ClassC{} has three positions: a leased in-process view, a supervised worker fed by a shared ring, and a portable E3 client over SCTP. Observations re-enter control only through host-owned caches and authorities, and one embedded E3 agent manages every class off the hot path.}
\label{fig:system}
\end{figure*}

\section{Design Consequences: The Released ABI}
\label{sec:design}

Each mechanism in the released runtime~\cite{pennybacker2026dapparch} can now be stated as a consequence of a measurement in Section~\ref{sec:meas} and a row class in Section~\ref{sec:corpus}. Table~\ref{tab:mechanisms} summarizes; the text gives the reasoning.

\begin{table*}[t]
\centering
\caption{Mechanism adopted per class, the measurement that justifies it, and what the alternative would have cost on the same host. Costs are P99.9 unless stated.}
\label{tab:mechanisms}
\scriptsize
\setlength{\tabcolsep}{3pt}
\begin{tabularx}{\textwidth}{@{}L{1.5cm}L{4.1cm}YL{4.3cm}@{}}
\toprule
\textbf{Class} & \textbf{Adopted mechanism} & \textbf{Measured cost of the adopted mechanism} & \textbf{Cheapest rejected alternative, and why} \\
\midrule
\textcolor{clsA}{A} & Direct C ABI on the lane's CUDA stream; device pointers in, DU-owned buffers out; CUDA-event completion; grant shape checked against the declared admission profile & \MechAStream{}; receiver at the live shape \ChkARxLive{} (P50 / P99.9) & Device-to-host export: \MechADtoH{} P99 at 1\,MB before inference, plus a wake and a return copy no framework provides \\
\textcolor{clsB}{B} & Direct call at the decision boundary; intents range-checked and allow-flagged; deadline in the request, checked at commit; circuit breaker after 8 failures & \MechBAbi{} over 20{,}000 validated calls (quiet), \ContBAbi{} with the DU on the air; 3.6\,\us{} on the runtime & Same request over SCTP loopback: \MechBSctp{} quiet, \ContBSctp{} with the DU on the air; ZeroMQ \code{ipc://}: \MechBZmq{} and \ContBZmq{} \\
\textcolor{clsC}{C} native & Leased zero-copy views; latest-context cache with 3 bounded acquires & Idle stream: 1 atomic load, \MechCIdle{}; context acquire \MechCCtx{} & A copy per subscriber (the ring's \MechCRingProd{}), paid only when isolation is wanted \\
\textcolor{clsC}{C} supervised & Shared ring: try-lock, reclaim read, 1 memcpy, release store, eventfd; full ring is a counted drop; SEQPACKET control with credentials; CUDA-IPC pool for device payloads & Producer at 68\,KB: \MechCRingProd{}, full-ring return \MechCRingSlow{}; consumer wake \MechCWake{}; control round trip \MechCSeq{} & ZeroMQ PUB/SUB at 68\,KB: producer \MechCZmqProd{}; a slow subscriber's data ages to \ZmqAgeSlow{}; conflation drops silently \\
\textcolor{clsC}{C} portable & E3AP over SCTP, APER via asn1c; FlatBuffers data profile on a second SCTP association; credentialed local socket & Control round trip \MechCERT{}; 64\,KB indication \MechCEInd{} one way; 1.47\,MB grid \MechCEGrid{} one way & None rejected: kept because 17 of 39 rows fit it and E2 tooling exists for it \\
\bottomrule
\end{tabularx}
\end{table*}

\subsection{\ClassA{}: A Stream, Not a Message}
The inline rows have 50--300\,\us{} budgets over 68\,KB to several MB of GPU-resident tensors, and Section~\ref{sec:meas:a} shows that any path leaving the PHY's stream exceeds the budget before a single model FLOP runs. The only admissible mechanism is enqueue on the PHY's own stream: the ABI passes a \code{cudaStream\_t} and device pointers, the DU records a completion event behind the dApp's work and keeps its conventional stage armed if the event has not fired by the deadline. Because the dApp is inside the process, the ABI is frozen and size-tagged, so a stale module is rejected at load, and the grant shape is checked against the module's declared admission profile before any module code runs.

\subsection{\ClassB{}: A Call With a Contract}
The bounded rows exchange kilobytes and wait for the answer, and Section~\ref{sec:meas:b} shows that every process boundary with a wake in it, however carried, moves the P99.9 from single microseconds to hundreds and adds misses at 100\,\us{} once the DU is on the air. A direct call has no such tail, so the ABI uses one and spends the saved budget on what a boundary would have provided: every intent is range-checked against the cell's PRB count, MCS table, and the operator's allow flags, the deadline travels in the request and is checked at commit, and a circuit breaker removes a module after eight recoverable failures. That suffices because the outputs are 72\,B each and fully validatable, and it is why the verified-hook alternative of Section~\ref{sec:meas:jbpf} was evaluated and not adopted, and why the supervised \ClassB{} position of Section~\ref{sec:meas:b}, which fits the budget only by spinning on a reserved core, is defined but not shipped.

\subsection{\ClassC{}: Three Positions, One Producer Rule}
The nonblocking rows tolerate milliseconds but their producer does not: the thread publishing a spectrum grid is on the slot's critical path. Section~\ref{sec:meas:c} shows that producer cost and accountable staleness, not consumer latency, are the constraints, and each \ClassC{} position follows from them. \emph{Leased views} cost the producer an atomic load when idle and a reference count when subscribed; the DU refuses to unload a module until its leases return, the quiescence proof of~\cite{pennybacker2026dapparch}. The \emph{supervised process} costs the producer a try-lock, a reclaim read, one memcpy, and a release store on a shared ring, plus one eventfd write to wake the worker. A full ring returns immediately as a counted drop, so staleness is bounded by the ring depth and every loss is visible. Control travels on a SEQPACKET socket with credentials verified per message, so a dead worker is detected within one heartbeat and restarted without touching the ring. GPU payloads use a CUDA-IPC pool. \emph{Portable E3} is the protocol-standard end: its measured costs (Section~\ref{sec:meas:e3}) are inside every \ClassC{} budget, so it is a legitimate choice for all 17 rows and the only choice outside the host's trust domain. The platform ships it with the same asn1c toolchain and message families as the Northeastern agent, plus a FlatBuffers profile for bulk streams on a second SCTP association with its own port.\footnote{The preview release's default port numbers for the two associations, 36423 and 38472, are IANA assignments to the 3GPP SLm and F1 control planes respectively, and the F1 collision matters on a host that also runs a CU. The ports are configuration keys and the defaults are being reassigned; the paper therefore refers to the associations by role rather than by number.}

\subsection{Closing the Loop Without a Message}
\label{sec:design:loop}
The reason \ClassC{} needs in-process positions at all is not speed, since Section~\ref{sec:meas:c} shows the ring and the socket cost the producer the same. It is the return path. An observation made by a \ClassC{} dApp re-enters control in one of two ways. Through the protocol-standard end it is an indication out and a control action back, \DrvESrs{} one way for a 64\,KB body on the quiet host and \ContEOneWaySrs{} with the DU on the air, plus \ContERT{} for the action, and the action is applied at a safe point after the slot that produced the observation. Through the latest-context cache it is a publish into a host-owned structure and a bounded acquire by the next \ClassB{} call, \MechCCtx{}, inside that call's 100\,\us{} budget. The interference map of the spectrum dApp reaching the scheduler's avoidance mask is the shipped instance, and it is the only row in the corpus in which an observation changes a per-slot decision with no message on the decision path. An observer-only framework can implement the observation and the control action; it cannot implement the cache, because the cache lives on the consuming side of the boundary.

Supporting three positions costs the platform three egress implementations behind one publish call and the dApp author nothing: the package declares its position, and the price of looser coupling (Table~\ref{tab:mechanisms}) is paid only by the dApp that chose it.

\section{Validation on the Released Platform}
\label{sec:validation}

The mechanisms of Section~\ref{sec:design} ship in the public \ocudu{} dApp platform and SDK~\cite{ocudu_dapp_platform,ocudu_dapp_sdk}. Table~\ref{tab:checkpoints} places the runtime's own per-class checkpoints beside this study's measurement of the same mechanism on the same host, so the microbenchmark and the production number can be compared directly; the reference receiver's earlier 92.5\,\us{} figure is reproduced at the live shape and, as Section~\ref{sec:meas:a} shows, does not yet extend to the 273-PRB kernels. The composition run in the last row is the practical test of the argument: four dApps from three classes, built out of tree against the installed SDK, loaded through one E3 agent onto one over-the-air n78 cell (GB10 host, USRP B210, 51\,PRB, 30\,kHz TDD) with three live UEs. The \ClassA{} equalizer replaced the conventional stage for every PUSCH grant, the \ClassB{} scheduler returned validated intents at every decision boundary, the \ClassC{} spectrum and SRS-ISAC dApps published at producer rate, one in-process and one as a supervised CUDA worker through the CUDA-IPC pool, and a Python E3 client subscribed to the same cell over SCTP from outside the process. The spectrum dApp's interference map reached the scheduler through the latest-context cache, the observation-to-control loop of Section~\ref{sec:design:loop} that no observer-only boundary closes inside the slot.

\begin{table}[t]
\centering
\caption{Runtime checkpoints and the corresponding suite measurement. Runtime column: the platform's own benchmark binaries run on this host for this study, except the 92.5 / 105.1\,\us{}, 45--52\,\us{}, 3.6\,\us{}, and 0.87\,Gb/s figures, which are from the release qualification. Suite column: this study, cell stopped, GPU exclusive.}
\label{tab:checkpoints}
\footnotesize
\setlength{\tabcolsep}{3pt}
\begin{tabularx}{\columnwidth}{@{}lYR{1.7cm}R{1.9cm}@{}}
\toprule
\textbf{Class} & \textbf{Mechanism} & \textbf{Runtime} & \textbf{Suite} \\
\midrule
A & Receiver completion at the live shape (51\,PRB, 2 layers, 256-QAM), P50 / P99.9 & 92.5 / 105.1\,\us{} & \ChkARxLive{} \\
A & Receiver completion at 273\,PRB, 2 layers, P50 / P99.9 (150\,\us{} deadline) & --- & \ChkARxFull{} \\
A & Estimator submit on the lane's stream, live shape, P50 / P99.9 & --- & \ChkAStream{} \\
A & Equalizer, live cell, P50 & 45--52\,\us{} & --- \\
B & Direct call P99.9 (runtime: worst profile, 9{,}000 calls; suite: 20{,}000 validated calls) & 3.6\,\us{} & \ChkBAbi{} \\
B & Latest-context acquire and release, P50 / P99.9 & \ChkBCtx{} & --- \\
C & Export across forced restarts; suite: ring producer per slot & 0.87\,Gb/s & \ChkCRing{} \\
C & E3AP round trip: runtime get-config (local socket) P50 / P99; suite: control action over SCTP & \ChkCERT{} & \ChkCERTsuite{} \\
\midrule
All & 4 dApps, 3 classes, 1 OTA cell & \multicolumn{2}{r@{}}{260{,}000+ inv., 0 fallbacks} \\
\bottomrule
\end{tabularx}
\end{table}

\section{Limitations}
\label{sec:limits}

All measurements come from one host, an NVIDIA DGX Spark with a GB10 GPU and coherent CPU--GPU memory. On a PCIe-attached GPU the device-to-host rows move by the bus latency and the \ClassA{} argument only strengthens; the host-side rows should transfer with the usual caveat that tails depend on kernel configuration and isolation. Both host conditions are archived as raw traces.

The DU-on-air condition is one cell at 51\,PRB with three UEs, and the benchmarks ran on six cores the DU does not use. That is a light load and a generous placement: a production supervised worker would share cores with the DU, more UEs would lengthen the DU's own bursts, and both move the message-path tails the wrong way. The contention figures are therefore a floor for the cost of a boundary, not an estimate of it.

The E3 rows measure the codec and the loopback carrier. A cross-host SCTP association adds network latency that depends entirely on the operator's fabric and is out of scope here. The 273-PRB/four-port rows are the same code at the envelope size and were not exercised over the air.

\ClassA{} is functionally resident and its interface cost is measured, but the reference receiver meets its deadline only at the live shape; the 273-PRB kernels are not yet qualified. Finally, the corpus verdicts are engineering judgments backed by byte and deadline arithmetic; a framework that adds an in-process position would move rows across the line, which is the point of the argument.

\section{Conclusion}
\label{sec:conclusion}
The dApp tier was born as an observer with a reply channel, and for the sensing and advisory use cases that motivated it that boundary remains right. The audit here shows that more than half of the AI-RAN use cases now filed under the same label cannot cross it, for reasons of arithmetic: a copy and a wake on a slot's critical path, or a process hop inside a 100\,\us{} decision, exceed the budget before the model runs, and the second of these is invisible on an idle host. Measuring the mechanisms the three public frameworks actually use prices each position on the coupling spectrum, and those prices derive the released ABI: a stream for \ClassA{}, a validated call for \ClassB{}, and for \ClassC{} a per-use-case choice among a lease, a supervised ring, and a protocol-standard E3 association.

Three recommendations follow for the groups now shaping the dApp definition. First, a dApp specification should carry the blocking contract and the payload residency as first-class attributes of a use case, because they decide the interface before any model does, and a definition that names only the observer contract excludes 22 of the 39 use cases the community has already proposed. Second, E3AP should remain the common management plane for every class, so that lifecycle, subscription, and control look the same to an operator whether the dApp is a shared object or a container, and so that existing E3 dApps run unchanged; the classes differ in their data path, not in how they are managed. Third, the in-process position needs an ABI that is frozen, size-tagged, admission-checked, and validated at the boundary, because that is what replaces the isolation a process would have given; the released \ocudu{} interface is offered as one candidate, and the benchmark suite as the way to price any other. The released platform composes all three classes on one cell and records the checkpoints that let the argument be re-run.

\section*{Availability and Acknowledgment}
The benchmark suite, corpus, raw CSVs, and this paper's sources accompany the released platform, SDK, and quickstart images, public under the \ocudu{} WG2 AI-RAN group~\cite{ocudu_dapp_platform,ocudu_dapp_sdk,ocudu_dapp_quickstart}. This work is supported by the U.S. Department of Defense (DoD) Office of the Under Secretary of Defense for Research and Engineering (OUSD(R\&E)) FutureG Office.


\clearpage
\onecolumn
\raggedbottom
\appendix[The Audited Use-Case Corpus]
\label{app:corpus}
\begin{table}[H]
\centering
\caption{Runtime rows of the audited corpus (identifiers are the ones used in the text). Bytes are per invocation at the live 51-PRB, 2-port cell and at the 273-PRB, 4-port envelope (64 ports for A-04). Verdicts for the observer-only boundary (E3 indication out, control action back) and the GPU-export boundary (copy into a second context): \emph{inexpr.}, no return path into the consuming stage; \emph{infeas.}, a path exists and its measured P99.9 with the DU on the air exceeds the budget (on the quiet host the same path fits); \emph{feas.}, fits; \emph{weak}, the row does not discriminate. Status is on the released platform.}
\label{tab:corpus_all}
\scriptsize
\setlength{\tabcolsep}{3pt}
\begin{tabularx}{\textwidth}{@{}lYlR{1.0cm}R{1.0cm}L{2.9cm}L{2.0cm}L{1.2cm}Y@{}}
\toprule
\textbf{ID} & \textbf{Use case} & \textbf{Class} & \textbf{Live} & \textbf{Max} & \textbf{Deadline / freshness} & \textbf{Observer-only} & \textbf{GPU export} & \textbf{\ocudu{} status} \\
\midrule
A-01 & Neural channel estimation & \textcolor{clsA}{A} & 67\,KB & 1.4\,MB & 100\,\us{} goal; 800\,\us{} wall & inexpr. & inexpr. & shipped (0x00010003) \\
A-02 & Neural equalization (CE included) & \textcolor{clsA}{A} & 67\,KB & 1.4\,MB & 100\,\us{} goal; 800\,\us{} wall & inexpr. & inexpr. & 45--52\,\us{} P50 live; 260k+ inv. \\
A-03 & Neural receiver to soft LLRs & \textcolor{clsA}{A} & 67\,KB & 1.4\,MB & 150\,\us{} goal; 2 ms wall & inexpr. & inexpr. & 92.5 / 105.1\,\us{} P50 / P99.9 \\
A-04 & Large-MIMO PUSCH at 64 rx ports & \textcolor{clsA}{A} & 11.7\,MB & 22.4\,MB & 1 UL slot occupancy & inexpr. & inexpr. & proxy 453\,\us{} P50, 1.4 ms P99 \\
A-05 & Joint neural EQ and FEC decoding & \textcolor{clsA}{A} & 176\,KB & 4.4\,MB & 300\,\us{}; soft buffer host-owned & inexpr. & inexpr. & out of scope (HARQ buffer) \\
A-06 & Neural PRACH detection & \textcolor{clsA}{A} & 13\,KB & 3.9\,MB & 1--2 ms per occasion & weak (rate) & weak & design only (no PRACH seam) \\
A-07 & Neural SRS channel estimation & \textcolor{clsA}{A} & 5\,KB & 210\,KB & SRS period 5--40 ms & weak (size) & weak & design only as A; C path shipped \\
A-08 & Inline neural precoding & \textcolor{clsA}{A} & 68\,KB & 2.1\,MB & DL slot; 100\,\us{} goal & inexpr. & inexpr. & out of scope (no DL seam) \\
A-09 & PAPR reduction and neural DPD & \textcolor{clsA}{A} & 60\,KB & 240\,KB & per symbol (35.7\,\us{}) & inexpr. & inexpr. & RU-seam plugin path (vendor id) \\
A-10 & ISAC waveform and beam writeback & \textcolor{clsA}{A} & 68\,KB & 2.1\,MB & DL slot & inexpr. & inexpr. & out of scope (sensing half is C-02) \\
A-11 & Two-sided learned modem & \textcolor{clsA}{A} & 68\,KB & 1.4\,MB & 50--300\,\us{} per allocation & inexpr. & inexpr. & out of scope (UE pairing) \\
A-12 & OTFS and Zak-OTFS air interface & \textcolor{clsA}{A} & 68\,KB & 1.4\,MB & 100--500\,\us{} per allocation & inexpr. & inexpr. & out of scope \\
A-13 & Semantic joint source-channel coding & \textcolor{clsA}{A} & 68\,KB & 1.4\,MB & 50--300\,\us{} per allocation & inexpr. & inexpr. & out of scope (DL seam; paired UE model) \\
\addlinespace[3pt]
B-01 & Scheduler intents & \textcolor{clsB}{B} & 512\,B & 4\,KB & 100\,\us{} to commit & infeas.\ (on air) & inexpr. & 3.6\,\us{} P99.9 runtime; 5.2\,\us{} suite \\
B-02 & Uplink power control (TPC) & \textcolor{clsB}{B} & 64\,B & 64\,B & 100\,\us{} (shared) & infeas. & inexpr. & shipped (TPC field) \\
B-03 & Link adaptation and MCS & \textcolor{clsB}{B} & 512\,B & 4\,KB & 100\,\us{} (shared) & infeas. & inexpr. & shipped (OLLA reference) \\
B-04 & Interference-aware PRB avoidance & \textcolor{clsB}{B} & 275\,B & 275\,B & 100\,\us{} to commit & infeas. & inexpr. & demonstrated live (K-01) \\
B-05 & Per-slot slice enforcement & \textcolor{clsB}{B} & 512\,B & 4\,KB & 100\,\us{} (shared) & infeas. & inexpr. & shipped (fields + envelope) \\
B-06 & Subband mute for SBFD / dynamic TDD & \textcolor{clsB}{B} & 275\,B & 275\,B & 100\,\us{} (shared) & infeas. & inexpr. & mask path shipped; host support open \\
B-07 & Beam selection at the decision boundary & \textcolor{clsB}{B} & 10\,KB & 10\,KB & 100\,\us{} (shared) & infeas. & inexpr. & output defined; input context absent \\
B-08 & Security-gated forbid & \textcolor{clsB}{B} & 64\,B & 64\,B & 100\,\us{} (shared) & infeas. & inexpr. & design only \\
B-09 & Layer reduction for energy & \textcolor{clsB}{B} & 512\,B & 4\,KB & 100\,\us{} (layers) & infeas. & inexpr. & nof\_layers intent; gain via C-16 \\
\addlinespace[3pt]
C-01 & Spectrum sensing and interference map & \textcolor{clsC}{C} & 134\,KB & 1.4\,MB & 0.5--10 ms; per captured slot & feas.\ (decimated) & feas. & 3 positions; 0.87 Gb/s across restarts \\
C-02 & Quiet-period reservation / DSS windows & \textcolor{clsC}{C} & 100\,B & 100\,B & $\ge$ 1 slot ahead & feas.\ (no admission) & inexpr. & shipped (reservation authority) \\
C-03 & SRS-ISAC feature extraction & \textcolor{clsC}{C} & 5\,KB & 210\,KB & $\le$ SRS period & feas. & feas. & live (CUDA-IPC pool) \\
C-04 & CSI prediction & \textcolor{clsC}{C} & 5\,KB & 210\,KB & 1--10 ms; $\ge$ 5 ms over E3 & feas. & feas. & data path shipped; consumer design \\
C-05 & Advisory beam prediction & \textcolor{clsC}{C} & 10\,KB & 10\,KB & $\le$ 1 slot native; 5--20 ms E3 & feas. & feas. & design only \\
C-06 & Beam failure prediction & \textcolor{clsC}{C} & 1\,KB & 1\,KB & 10--200 ms timer & feas. & feas. & design only \\
C-07 & Positioning and channel charting & \textcolor{clsC}{C} & 5\,KB & 210\,KB & 10 ms--1 s & feas. & feas. & design only (srs.channel) \\
C-08 & Positioning measurement enhancement & \textcolor{clsC}{C} & 5\,KB & 210\,KB & 1--20 ms & feas. & feas. & design only \\
C-09 & Joint sensing and positioning & \textcolor{clsC}{C} & 5\,KB & 210\,KB & 10 ms--1 s & feas. & feas. & design only \\
C-10 & RF fingerprinting & \textcolor{clsC}{C} & 134\,KB & 2.9\,MB & 0.5--10 ms & feas. & feas. & design only (pusch.grid) \\
C-11 & Physical-layer security detection & \textcolor{clsC}{C} & 134\,KB & 1.4\,MB & 0.5--10 ms; gating via B-08 & feas. & feas. & design only \\
C-12 & Codebook and beam-weight optimisation & \textcolor{clsC}{C} & 5\,KB & 210\,KB & 10 ms--1 s & feas. & feas. & design only (xApp candidate) \\
C-13 & CSI feedback reconstruction & \textcolor{clsC}{C} & 1\,KB & 1\,KB & $\le$ 1 slot fresh; 2 ms cadence & weak & weak & design only (no UCI seam) \\
C-14 & Inter-cell interference coordination advisory & \textcolor{clsC}{C} & 275\,B & 275\,B & 1--10 ms; multi-cell via E2 & feas. & feas. & design only \\
C-15 & RIS advisory & \textcolor{clsC}{C} & 100\,KB & 100\,KB & 1--20 ms & feas. & feas. & out of scope \\
C-16 & RF gain policy & \textcolor{clsC}{C} & 1\,KB & 1\,KB & 1--10 ms actuation & feas. & feas. & actuation shipped (ranControl) \\
C-17 & Portable external observer & \textcolor{clsC}{C} & 134\,KB & 1.4\,MB & stream floors & feas.\ (native) & feas. & live (Python E3 client) \\
\bottomrule
\end{tabularx}
\end{table}

Table~\ref{tab:corpus_all} lists the 39 runtime rows and Table~\ref{tab:corpus_other} the xApp/rApp and research rows that Section~\ref{sec:corpus} excludes from the runtime count; the identifiers are the ones used throughout the paper, and the mapping to the earlier corpus of~\cite{pennybacker2026dappclasses} is kept in the repository. The five compositions counted in Section~\ref{sec:corpus} couple rows already listed here through host-owned caches and authorities, so they are described in the study repository rather than given separate entries. The byte counts are the inputs each row needs per invocation, computed from the 5G NR grid dimensions at 30\,kHz subcarrier spacing; the deadline column gives the 5G NR bound or, for asynchronous rows, the freshness after which the result loses value. The verdict columns apply the arithmetic of Section~\ref{sec:meas} to the observer-only boundary (an E3 indication out and a control action back) and to a GPU-export boundary (device-to-host or device-to-device copy into a second context). The status column names what the released platform ships, has demonstrated on the air, defines but has not implemented, or leaves out of scope. The same rows, with their formulas and sources, are in the study repository's use-case inventory and performance matrix.

\begin{table}[H]
\centering
\caption{Rows that are not dApps: management-plane applications acting through typed configuration or RAN control, xApp/rApp policy, and one research row that needs a multi-DU substrate. They consume summaries or policies and do not size the microsecond interfaces.}
\label{tab:corpus_other}
\scriptsize
\setlength{\tabcolsep}{3pt}
\begin{tabularx}{\textwidth}{@{}lYL{1.8cm}L{2.6cm}Y@{}}
\toprule
\textbf{ID} & \textbf{Application} & \textbf{Bucket} & \textbf{Timescale} & \textbf{\ocudu{} status} \\
\midrule
X-01 & QoS and slice-aware flow classification & xApp/rApp & 1--10 ms & out of scope: no packet-path seam \\
X-02 & RAN fault and anomaly detection & xApp/rApp & 10 ms--1 s & live: MCP observe ring; ranControl actuation \\
X-03 & Energy saving, carrier sleep, DTX/DRX & xApp/rApp & 100 ms--1 s & partial: actuation shipped; policy out of scope \\
X-04 & SON self-optimisation & xApp/rApp & > 1 s & out of scope: out of scope \\
X-05 & Slice SLA management & xApp/rApp & > 1 s & out of scope: out of scope \\
X-06 & Handover and load-balancing advisory & xApp/rApp & 10 ms--1 s & partial: handoverUe / releaseUeToIdle shipped \\
X-07 & Fronthaul and UL-throughput configuration & mgmt. & 10 ms--1 s & partial: gain actuation shipped \\
X-08 & Slice envelope & mgmt. & 100 ms--1 s & shipped: typed configuration writes \\
X-09 & LLM-agent operations (28 MCP tools) & mgmt. & seconds & live: MCP server on the E3AP endpoint \\
R-02 & Distributed and cell-free MIMO & research & per slot / 100 ms & out of scope: multi-DU substrate \\
\bottomrule
\end{tabularx}
\end{table}

\end{document}